\pdfoutput=1
\documentclass[traditabstract]{aa}

\usepackage[squaren,Gray]{SIunits}

\usepackage{subfig}
\usepackage{graphicx}
\usepackage{txfonts}
\usepackage{empheq}
\usepackage{float}

\usepackage{natbib}
\bibpunct{(}{)}{;}{a}{}{,}

\usepackage{hyperref}
\hypersetup{colorlinks=true,citecolor=blue,linkcolor=blue}

\newcommand{\vb}[1]{\vec{#1}}
\newcommand{\norm}[1]{\Vert#1\Vert}
\newcommand{\dvt}[1]{\frac{d#1}{dt}}

\newcommand{\AUd}{\mathrm{au}^3/\mathrm{day}^{2}}

\newcommand{\arc}{\arcsecond\per\mathrm{yr}}

\renewcommand{\b}{\beta}
\renewcommand{\d}{\delta}
\renewcommand{\dvt}{\dot}

\renewcommand{\sun}{0}
\newcommand{\planet}{1}
\newcommand{\orbit}{1}

\newcommand{\e}{\varepsilon}
\renewcommand{\H}{\mathcal{H}}
\newcommand{\R}{\mathcal{R}}
\newcommand{\G}{\mathcal{G}}
\newcommand{\lp}{\lambda_\orbit-\psi}
\newcommand{\CR}{\mathcal{E}}
\newcommand{\fe}{\left(1+\frac{3}{2} e^2\right)}
\newcommand{\Dlim}{\sigma_{s}}
\newcommand{\npr}{n'_1}
\newcommand{\np}{n_1}
\newcommand{\Alib}{A_{lib}}
\newcommand{\Atot}{A_{tot}}
\newcommand{\dAlib}{\dot{A}_{lib}}
\newcommand{\dAtot}{\dot{A}_{tot}}

\newcommand{\ur}{\vb{\hat r}_\orbit}
\newcommand{\vecc}{\vb{e}}
\newcommand{\necc}{\vb{u}}
\newcommand{\vrot}{\vb{S}}
\newcommand{\nrot}{\vb{s}}
\newcommand{\norb}{\vb{k}}
\newcommand{\peri}{\vb{i}}
\def\crm{\cr\noalign{\medskip}}

\begin{document}

\title{Eviction-like resonances for satellite orbits}
\subtitle{Application to Phobos, the main satellite of Mars}
\titlerunning{Eviction-like resonances for satellite orbits}

\author{
Timoth\'ee Vaillant\inst{1,2}
 \and
Alexandre C. M. Correia\inst{2,3}
}

\institute{
CIDMA, Departamento de Matem\'atica, Universidade de Aveiro, Campus de Santiago, 3810-193 Aveiro, Portugal
\and
CFisUC, Departamento de F\'isica, Universidade de Coimbra, 3004-516 Coimbra, Portugal
\and 
ASD, IMCCE-CNRS UMR8028, Observatoire de Paris, PSL Universit\'e, 77 Av. Denfert-Rochereau, 75014 Paris, France
}


\abstract{The motion of a satellite can experience secular resonances between the precession frequencies of its orbit and the mean motion of the host planet around the star.
Some of these resonances can significantly modify the eccentricity (evection resonance) and the inclination (eviction resonance) of the satellite.
In this paper, we study in detail the secular resonances that can disturb the orbit of a satellite, in particular the eviction-like ones.
Although the inclination is always disturbed while crossing one eviction-like resonance, capture can only occur when the semi-major axis is decreasing.
This is, for instance, the case of Phobos, the largest satellite of Mars, that will cross some of these resonances in the future because its orbit is shrinking owing to tidal effects. 
We estimate the impact of resonance crossing in the orbit of the satellite, including the capture probabilities, as a function of several parameters, such as the eccentricity and the inclination of the satellite, and the obliquity of the planet.
Finally, we use the method of the frequency map analysis to study the resonant dynamics based on stability maps, and we show that some of the secular resonances may overlap, which leads to chaotic motion for the inclination of the satellite.}


\maketitle

\section{Introduction}

The orbit of a satellite can give some hints as to its origin.
If the satellite forms at the same time as its planet within an accretion disk, it is expected that the satellite is located in the equatorial plane of the planet in a nearly circular orbit \citep[e.g.,][]{Batygin_Morbidelli_2020, Inderbitzi_etal_2020}.
If the satellite forms from a giant impact, the initial orbit is also expected to be equatorial, but it can be very eccentric \citep[e.g.,][]{Canup_Asphaug_2001, Canup_2005}.
Finally, if the satellite is a captured object, its orbit is expected to be very eccentric as well and to present some inclination with respect to the planet's equator  \citep[e.g.,][]{Agnor_Hamilton_2006, Nesvorny_etal_2007}.

Tidal dissipation inside the planet and the satellite can significantly modify some orbital parameters, in particular the semi-major axis and the eccentricity.
Tidal dissipation can also lead to variations in the inclination of the satellite, although these variations are weak and often negligible \citep[e.g.,][]{lambeck1979, szeto1983}.
Therefore, the inclination is not modified by tides much and, contrary to the semi-major axis and the eccentricity, it can be used to distinguish between the capture scenario and other formation scenarios.
Nevertheless, secular resonances can also modify the orbital parameters of the satellite.
For instance, in the case of Phobos, the main satellite of Mars, \cite{yoder1982} noted that during its evolution, the satellite crossed several resonances which have successively excited its eccentricity.
Therefore, the current value cannot be considered simply as the result of the tidal dissipation of the primordial eccentricity.
The evection resonance, which corresponds to an interaction between the secular precession of the pericenter of the satellite and the orbital mean motion of the planet, can also lead to a variation in the eccentricity of the satellite \citep[e.g.,][]{touma1998}.
This effect is stronger when capture in the evection resonance occurs, although this is only possible if the semi-major axis is increasing, which is, for instance, the case for the early Moon.

There are also resonances that are able to modify the inclination of a satellite, corresponding to an interaction between the precession of the node of its orbit and the orbital mean motion of the planet.
\citet{touma1998} noted that the early Moon could also have been temporarily captured in a resonance of this kind, which they called "eviction" (just after it escapes from the evection resonance).
This temporary capture can lead to an increase of around ten degrees in the inclination of the early Moon with respect to the Earth's equator, which may explain the current mutual inclination of the lunar orbit of about five degrees.
The capture of the Moon in the eviction resonance is only possible if the semi-major axis is decreasing and if the eccentricity is sufficiently high.
\cite{touma1998} also noted that if the Moon encounters the eviction resonance while the semi-major axis increases, it would not be captured, but it could nevertheless lead to a one-time increase in the inclination, provided that the eccentricity is high enough.

As the semi-major axis of Phobos is currently decreasing due to tides \citep[e.g.,][]{jacobson2014}, it can also be captured in eviction-like resonances.
Indeed, \cite{yokoyama2002} showed that Phobos will encounter two of these resonances in the near future, which may significantly increase its equatorial inclination.
Contrary to the original eviction resonance described by \citet{touma1998}, capture in these two resonances is possible for zero eccentricity.
In order to study these resonances, \cite{yokoyama2002} considered a system of three bodies with the Sun, Mars, and a massless satellite, where the orbit of Mars is circular.
For this system, \cite{yokoyama2002} numerically integrated the equations of the satellite obtained from an averaged Hamiltonian, and noted that the capture probabilities in these resonances strongly depend on the Martian obliquity.
When capture does not occur, the resonance crossing can still lead to a sudden variation in the inclination.
The capture probability in one resonance is almost certain if the obliquity of Mars is higher than $20$ degrees.
Similar results for this resonance are observed in the case where the precession of the Martian equator, the obliquity variations \citep{yokoyama2002}, or the planetary perturbations on Mars are considered \citep{yokoyama2005}.
For the other resonance, which occurs almost simultaneously with the evection resonance, \cite{yokoyama2002} numerically observed that capture is possible with the resonant Hamiltonian, but becomes impossible with the total Hamiltonian.
\cite{yokoyama2002} then concluded that the impossibility of the capture is due to the interaction between the two resonances.
\cite{yokoyama2005} considered a more complete model including the Mars' eccentricity, and observed that it increases the interaction with the evection resonance and its effects.

In this paper, we revisit the secular resonances between the orbit of the satellite and the mean motion of the planet.
We focus on those that are suitable to excite the inclination of the satellite (eviction-like), since the inclination can be used to put constrains on the formation scenarios.
The aim of the paper is to estimate what can be the effects of such resonances on the orbital evolution of a satellite.
For this, we evaluate precisely the capture probabilities of a satellite in these resonances, and the variations in orbital parameters due to their crossings in the case where the capture fails.
We also study the interaction with the evection resonance observed by \cite{yokoyama2002} to determinate in which extent it influences the dynamics of a satellite and determine the exact mechanism behind this excitation.

In Sect.\,\ref{model}, we first derive a secular model that is suitable to describe the dynamics of a massless satellite of a rigid planet perturbed by the host star.
In Sect.\,\ref{resHam}, we identify all the possible secular resonances when the planet is in a circular orbit around the star and its equator uniformly precesses with constant obliquity.
In Sect.\,\ref{evicition}, we study in detail with analytical models the main eviction-like resonances that may significantly modify the inclination of the satellite.
In Sect.\,\ref{interaction}, we study the interaction of neighbor secular resonances using frequency map analysis.
In Sect.\,\ref{phobos}, we apply our model to the future evolution of Phobos and numerically estimate the capture probabilities in resonance for different values of the eccentricity and obliquity.
Finally, we discuss our results in Sect.\,\ref{concdisc}.

\section{Model}
\label{model}

We consider a star orbited by a planet and a satellite, with masses $m_\sun$, $m_\planet$ and $m$, respectively.
The star and the satellite are considered point masses. 
The planet is a rigid body with moments of inertia $A\leq  B\leq C$, and rotational angular momentum $\vrot$.

The variables ($a$, $e$, $i$, $\lambda$, $\varpi$, $\Omega$) denote respectively the semi-major axis, the eccentricity, the inclination with respect to a reference plane, the mean longitude ($\lambda=\varpi+M$, with $M$ the mean anomaly), the longitude of the pericenter ($\varpi=\omega+\Omega$, with $\omega$ the argument of the pericenter), and the longitude of the ascending node of the orbit of the satellite around the planet.
The same variables with a subscript $_\orbit$ are used for the orbit of the planet around the star, that we name for simplicity, the ecliptic.

\subsection{Secular Hamiltonian}

\begin{figure}[t]
\centering
\includegraphics[width=\columnwidth]{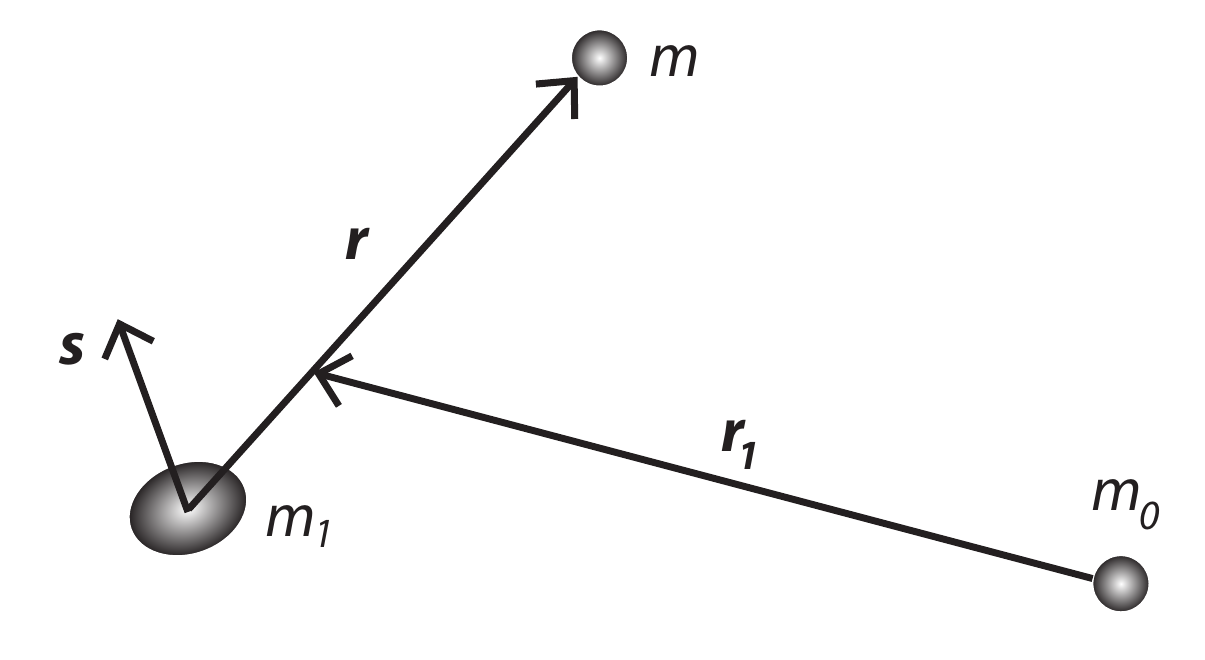}
\caption{Jacobi coordinates, where $\vb{r}$ is the position of $m$ relative to $m_\planet$ (satellite orbit), and $ \vb{r}_\orbit $ is the position of the center of mass of $m$ and $m_\planet$ relative to $m_\sun$ (planet orbit). The planet is a rigid body, where $\vrot$ is the rotational angular momentum.}
\label{fig1}
\end{figure}

We let $(\tilde{\vb{R}}_i,\vb{R}_i)$ be the barycentric canonical variables with $\vb{R}_i$ the position vector with respect to the barycenter of the system, and $\tilde{\vb{R}}_i=m_i\dot{\vb{R}}_i$ the conjugate momenta ($i=0,1$, for the star, and planet, respectively, and we do not use subscripts for the variables of the satellite).
We perform a canonical change into Jacobi variables $(\tilde{\vb{r}}_i,\vb{r}_i)$ as $\vb{r}_0=\vb{R}_0$, $\vb{r}_\orbit=\d\vb{R}+(1-\d)\vb{R}_1 -\vb{R}_0$, and $\vb{r}=\vb{R}-\vb{R}_1 $, with $\d=m/(m_\planet +m)$ (see Fig. \ref{fig1}).
In the quadrupolar three-body problem approximation and with an average on the fast angles of the rotation, the Hamiltonian of the system is given by \citep[e.g.,][]{Smart_1953,boue2006}
\begin{equation}
\begin{split}
\H= & \frac{\tilde{\vb{r}}_\orbit ^2}{2\b_\orbit }+\frac{\tilde{\vb{r}}^2}{2\b}-\frac{\G m_\sun m_\planet }{r_{01}}-\frac{\G m_\sun m}{r'}-\frac{\G m_\planet m}{r} \\
& -\frac{\CR m}{r^3}\left[1-3\left(\frac{\vb{r} \cdot\nrot}{r}\right)^2\right]-\frac{\CR m_\sun }{r_{01}^3}\left[1-3\left(\frac{\vb{r}_{01} \cdot\nrot}{r_{01}}\right)^2\right] \ ,
\end{split}
\end{equation}
where $\G$ is the gravitational constant, $\beta = m_\planet  m/(m_\planet +m)$, $\beta_\orbit = m_\sun  (m_\planet  + m)/(m_\sun +m_\planet +m)$, 
$\vb{r}_{01} = \vb{R}_1  - \vb{R}_0=\vb{r}_\orbit-\d\vb{r}$,
$\vb{r}' = \vb{R} - \vb{R}_0=\vb{r}_\orbit+(1-\d)\vb{r}$, $r=\norm{\vb{r}}$, $r'=\norm{\vb{r}'}$, $r_{01}=\norm{\vb{r}_{01}}$, $\nrot=\vrot/\norm{\vrot}$, and
\begin{equation}
\CR=\frac{\G}{2} \left(C-\frac{A+B}{2} \right) \left(1-\frac32 \sin^2J\right) \ .
\end{equation}
The angle $J$ between the rotational angular momentum and the axis of maximal inertia of the planet is very weak for the planets of the solar system, and so we assume $\sin J=0$.
We proceed to a development to the degree one in $\d$ and to the degree two in $r/r_\orbit$.
The Hamiltonian then becomes
\begin{equation}
\begin{split}
\H = & \frac{\tilde{\vb{r}}_\orbit ^2}{2\b_\orbit }+\frac{\tilde{\vb{r}}^2}{2\b}-\frac{\mu_\orbit \b_\orbit }{r_\orbit}-\frac{\mu\b}{r}+\frac{\G \b m_\sun}{2r_\orbit^3}\left[r^2-\frac{3\left(\vb{r}_\orbit  \cdot\vb{r}\right)^2}{r_\orbit^2}\right] \\
& -\frac{\CR m}{r^3}\left[1-3\left(\frac{\vb{r} \cdot\nrot}{r}\right)^2\right] -\frac{\CR m_\sun }{r_\orbit^3}\left[1-3\left(\frac{\vb{r}_{1} \cdot\nrot}{r_{1}}\right)^2\right] \\
& -3\d\frac{\CR m_\sun }{r_\orbit^3}\left[\frac{\vb{r}_\orbit  \cdot\vb{r}}{r_\orbit^2}-5\frac{\left(\vb{r}_\orbit  \cdot\vb{r}\right)\left(\vb{r}_\orbit  \cdot\nrot\right)^2}{r_\orbit^4}+2\frac{\left(\vb{r}_\orbit  \cdot\nrot\right)\left(\vb{r} \cdot\nrot\right)}{r_\orbit^2}\right] \ ,
\end{split}
\end{equation}
with $\mu=\mathcal{G}(m_\planet +m)$ and $\mu_\orbit =\mathcal{G}(m_\sun +m_\planet +m)$.

We now consider that the satellite is a massless particle.
We thus perform a canonical change of variables $(\tilde{\vb{r}},\vb{r})\rightarrow(\tilde{\vb{r}}',\vb{r})$, with $\tilde{\vb{r}}'=\tilde{\vb{r}}/m$, and then make $m \rightarrow 0$, in order to obtain the Hamiltonian for the motion of the perturbed satellite
\begin{equation}
\begin{split}
\H = & \frac{\tilde{\vb{r}}'^2}{2}-\frac{\mu}{r}+\frac{\G m_\sun }{2r_\orbit^3}\left[r^2-\frac{3\left(\vb{r}_\orbit  \cdot\vb{r}\right)^2}{r_\orbit^2}\right]-\frac{\CR}{r^3}\left[1-3\left(\frac{\vb{r} \cdot\nrot}{r}\right)^2\right] \\
& -\frac{3\CR}{r_\orbit^3}\frac{m_\sun }{m_\planet }\left[\frac{\vb{r}_\orbit  \cdot\vb{r}}{r_\orbit^2}-5\frac{\left(\vb{r}_\orbit  \cdot\vb{r}\right)\left(\vb{r}_\orbit  \cdot\nrot\right)^2}{r_\orbit^4}+2\frac{\left(\vb{r}_\orbit  \cdot\nrot\right)\left(\vb{r} \cdot\nrot\right)}{r_\orbit^2}\right].
\end{split}
\end{equation}

Finally, we average the orbit of the satellite over the mean anomaly \citep[for details see][]{boue2006}, and obtain the secular Hamiltonian 
\begin{equation}
\label{eq:ham1}
\begin{split}
\H_s = & \frac{3\G m_\sun a^2}{2r_\orbit^3}\left[\vecc^2+\frac{1}{2}\frac{\left(\vb{r}_\orbit  \cdot\necc\right)^2}{r_\orbit^2}-\frac{5}{2}\frac{\left(\vb{r}_\orbit  \cdot\vecc\right)^2}{r_\orbit^2}\right] \\
& +\frac{\CR}{2a^3 \norm{\necc}^5} \left[\necc^2-3\left(\nrot \cdot \necc\right)^2\right]\\
&+ \frac{9 a \CR }{r_\orbit^3}\frac{m_\sun }{m_\planet }\left[\frac{\vb{r}_\orbit  \cdot\vecc}{2 r_\orbit^2}\left(1-5\frac{\left(\vb{r}_\orbit  \cdot\nrot\right)^2}{r_\orbit^2}\right)+\frac{\left(\vb{r}_\orbit  \cdot\nrot\right)\left(\vecc \cdot \nrot\right)}{r_\orbit^2}\right] \ ,
\end{split}
\end{equation}
with
\begin{equation}
\vecc=e\,\peri \ , \quad \mathrm{and} \quad \necc=\sqrt{1-e^2}\,\norb \ ,
\end{equation}
where $\norb=\vb{G}/G$ is the normal to the orbit of the satellite, $\peri$ is the unit vector indicating the direction of its pericenter, $\vb{G}$ is the orbital angular momentum, $G=L\sqrt{1-e^2}$, and $L=\sqrt{\G m_\planet a}$.
The first term of $\H_s$ (Eq.\,(\ref{eq:ham1})) corresponds to the  perturbations from the central star, the second term corresponds to perturbations from the gravitational flattening of the planet, while
the third term describes the effect of the couple exerted by the central star on the rigid planet on the orbital motion of the satellite.

\begin{figure}
\centering
\includegraphics[scale=0.4]{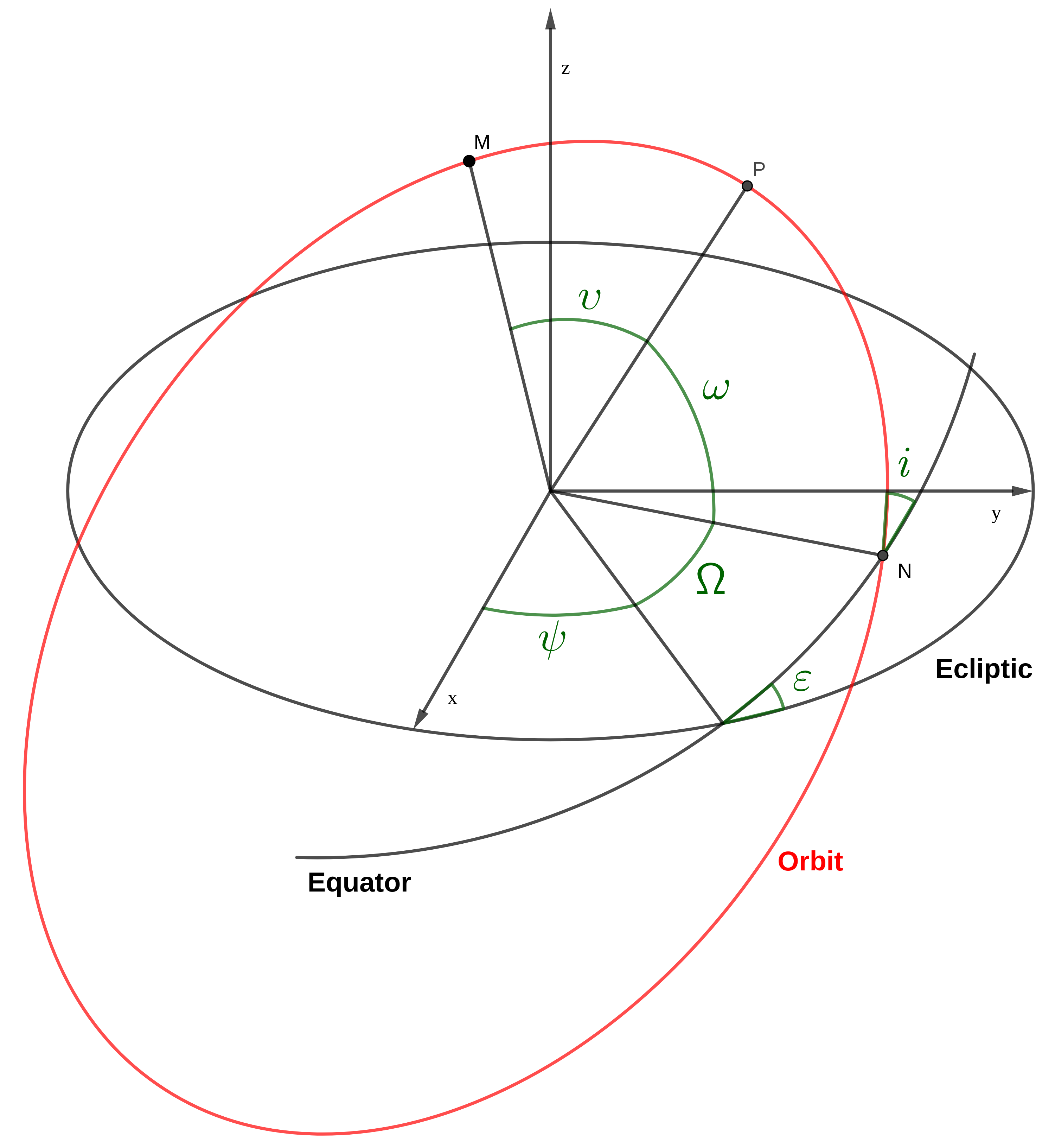}
\caption{Orbital elements of the orbit of the satellite (in red).}
\label{fig:orbit}
\end{figure}

\subsection{Equations of motion}

The equations of motion for the vectors $\vecc$ and $\necc$ can be obtained from the secular Hamiltonian $\H_s$ (Eq.\,(\ref{eq:ham1})) as \citep{tremaine2009, farago2010}
\begin{equation}
\begin{cases}
\dfrac{d \vecc}{dt} =\dfrac{1}{L}\left(\nabla_{\necc} \H_s \times\vecc+\nabla_{\vecc} \H_s \times\necc\right) \ , \crm
\dfrac{d \necc}{dt} =\dfrac{1}{L}\left(\nabla_{\necc} \H_s \times\necc+\nabla_{\vecc} \H_s \times\vecc\right) \ ,
\end{cases}
\label{eq:num1} 
\end{equation}
which gives
\begin{equation}
\begin{cases}
\dfrac{d \vecc}{dt} =\vb{h}_1\times\vecc+\vb{h}_2\times\necc+\gamma \, \necc\times\vecc \ , \crm
\dfrac{d \necc}{dt} =\vb{h}_1\times\necc+\vb{h}_2\times\vecc  \ ,
\end{cases}
\label{eq:num2} 
\end{equation}
with
\begin{equation}
\begin{split}
\vb{h}_1= &\frac{1}{L}\left[\frac{3\G m_\sun a^2}{2r_\orbit^5}\left(\vb{r}_\orbit \cdot\necc\right)\vb{r}_\orbit -\frac{3\CR}{a^3\norm{\necc}^5}\left(\nrot\cdot\necc\right)\vb{\nrot}\right] \ , \nonumber
\end{split}
\end{equation}
\begin{equation}
\begin{split}
\vb{h}_2= & \frac{1}{L}\left[-\frac{15\G m_\sun a^2}{2r_\orbit^5}\left(\vb{r}_\orbit \cdot\vecc\right)\vb{r}_\orbit \right. \\
&\left. +\frac{9\CR am_\sun }{r_\orbit^5 m_\planet }\left(\frac{\vb{r}_\orbit}{2}-\frac{5}{2}\frac{\left(\vb{r}_\orbit \cdot\nrot\right)^2}{r_\orbit^2}\vb{\vb{r}_\orbit } + \left(\vb{r}_\orbit \cdot\nrot\right)\nrot \right) \right] \ ,\nonumber
\end{split}
\end{equation}
\begin{equation}
\gamma=-\frac{1}{L}\left[\frac{3\CR}{2a^3\norm{\necc}^5}\left(1-5 \left(\nrot\cdot\norb\right)^2\right)+\frac{3\G m_\sun a^2}{r_\orbit^3}\right] \ . \nonumber
\end{equation}
We use Eq. (\ref{eq:num2}) to obtain numerically the secular orbital evolution of the satellite.
To completely solve these equations, we also need to compute the motion of the planet around the star, $\vb{r}_\orbit (t)$, and the evolution of the spin axis, $\nrot(t)$.
For simplicity, we assume that the planet moves on a circular orbit and that most of the angular momentum is on its orbit.
We also assume that the spin of the planet can only precess around the normal to the ecliptic with constant speed and obliquity (i.e., the angle between the equator of the planet and the ecliptic).
Therefore, we have
\begin{equation}
\vb{r}_\orbit \left(t\right)=\R_3\left(n_\orbit t \right)\begin{pmatrix}
a_\orbit \\ 0 \\ 0 \\
\end{pmatrix}  \ ,
\end{equation}
and
\begin{equation}
\nrot\left(t\right)=\R_3\left(\dot \psi t +\psi\left(0\right)\right) 
\R_1 \left(\varepsilon\right)\begin{pmatrix}
0 \\ 0 \\ 1 \\
\end{pmatrix} \ ,
\end{equation}
where $n_\orbit$ is the mean motion, $\e$ is the obliquity, and $\psi$ is the precession angle, which measures the precession of the equator with respect to the ecliptic (Fig.\,\ref{fig:orbit}). 
$\R_1$ and $\R_3$ are the rotations about the $x-$axis and $z-$axis, respectively
\begin{equation}
\R_1\left(\alpha\right)=\begin{pmatrix}
1 & 0 & 0 \\
0 & \cos\alpha & -\sin\alpha \\
0 & \sin\alpha & \cos\alpha \\
\end{pmatrix} \ ,
\end{equation}
\begin{equation}
\R_3\left(\alpha\right)=\begin{pmatrix}
\cos\alpha & -\sin\alpha & 0 \\
\sin\alpha & \cos\alpha & 0 \\
0 & 0 & 1 \\
\end{pmatrix} \ .
\end{equation}

\section{Secular resonances}
\label{resHam}

In general, the last term of the secular Hamiltonian (Eq.\,(\ref{eq:ham1})) is much weaker than the other two terms.
For instance, in the case of the Martian satellite Phobos, the ratio between the amplitudes of the third and first term is $\sim 10^{-9}$.
Moreover, on average, the effect of this third term is null over a satellite pericenter precession period.
Therefore, we can often neglect its contribution, and 
the secular Hamiltonian (Eq.\,(\ref{eq:ham1})) is rewritten as
\begin{equation}
\label{eq:hamSP}
\begin{split}
\H= & \frac{3\G m_\sun a^2}{2a_\orbit^3}\left[e^2+\frac{1-e^2}{2}\left(\ur \cdot\norb\right)^2-\frac{5e^2}{2}\left(\ur \cdot\peri\right)^2\right] \\
& +\frac{\CR}{2a^3 \left(1-e^2\right)^{3/2}} \left[1-3\left(\nrot \cdot \norb\right)^2\right] \ ,
\end{split}
\end{equation}
with $\ur=\vb{r}_\orbit /r_\orbit$.
Resonances between the orbit of the satellite and the Sun can only occur due to the first term.
In order to identify all possible resonances, we need to develop $\H$ in trigonometric series.
This development is given by \cite{yokoyama2002} but with some misprints.
A corrected version is presented in \cite{carvalho2020}.

We consider here the ecliptic as reference plane (Fig.\,\ref{fig:orbit}). 
Assuming that the satellite is a massless particle and that the orbital angular momentum of the planet is much larger than its rotational angular momentum, the ecliptic remains an inertial plane to a very good approximation.
We then need to express $(\ur \cdot\norb)^2$ and $(\ur \cdot\peri)^2$ as a function of the orbital elements of the satellite.
The unit vectors $\norb$, $\peri$ and $\ur$ can be expressed with respect to the ecliptic as a succession of rotations
\begin{equation}
\norb=\R_3\left(\psi\right) \R_1 \left(\e\right) \R_3\left(\Omega\right) \R_1 \left(i\right)\begin{pmatrix}
0 \\ 0 \\ 1 \\
\end{pmatrix} \ ,
\end{equation}
\begin{equation}
\peri=\R_3\left(\psi\right) \R_1\left(\e\right)
\R_3\left(\Omega\right)
\R_1\left(i\right)
\R_3\left(\omega\right)\begin{pmatrix}
1 \\ 0 \\ 0 \\
\end{pmatrix} \ ,
\end{equation}
and
\begin{equation}
\ur=\R_3\left(\lambda_\orbit\right)\begin{pmatrix}
1 \\ 0 \\ 0 \\ 
\end{pmatrix} \ ,
\end{equation}
where $\lambda_\orbit=n_\orbit t$ is the mean longitude of the planet (we assume a circular orbit).
The Hamiltonian (Eq.\,(\ref{eq:hamSP})) then becomes
\begin{equation}
\label{eq:hamdev}
\begin{split}
&\H=\frac{\CR}{2a^3 \left(1-e^2\right)^{3/2}} \left[1-3\cos^2 i\right]+\frac{3\G m_\sun a^2}{2a_\orbit^3}\left[-\frac{e^2}{4}\right.\\
& +\frac{1}{4}\fe\left(\sin^2i+\sin^2\e-\frac{3}{2}\sin^2i\sin^2\e\right)\\
&-\frac{1}{8}\fe\sin^2 i\sin^2\e\cos\left(2\Omega\right) \\
&+\frac{1}{8}\fe\sin\left(2i\right)\sin\left(2\e\right)\cos\left(\Omega\right)\\
&-\frac{5}{8}e^2\sin^2i\left(1-\frac{3}{2}\sin^2\e\right)\cos\left( 2\omega\right)\\
&-\frac{5}{32}e^2 \left(1+\cos i\right)^2\sin^2\e\cos\left( 2\left(\Omega+\omega\right)\right)\\
&-\frac{5}{32}e^2 \left(1-\cos i\right)^2\sin^2\e\cos\left( 2\left(\Omega-\omega\right)\right)\\
&-\frac{5}{16}e^2\sin i\left(1+\cos i\right)\sin\left(2\e\right)\cos\left( \Omega+2\omega\right)\\
&+\frac{5}{16}e^2\sin i\left(1-\cos i\right)\sin\left(2\e\right)\cos\left( \Omega-2\omega\right)\\
&-\frac{1}{4}\fe\left(1-\frac{3}{2}\sin^2 i\right)\sin^2\e\cos\left( 2\left(\lp\right)\right)\\
&-\frac{1}{16}\fe\sin^2 i\left(1-\cos \e\right)^2\cos\left( 2\left(\lp+\Omega\right)\right)\\
&-\frac{1}{16}\fe\sin^2 i\left(1+\cos \e\right)^2\cos\left( 2\left(\lp-\Omega\right)\right)\\
&+\frac{1}{8}\fe\sin\left(2i\right)\sin\e\left(1-\cos\e\right)\cos\left( 2\left(\lp\right)+\Omega\right)\\
&-\frac{1}{8}\fe\sin\left(2i\right)\sin\e\left(1+\cos\e\right)\cos\left( 2\left(\lp\right)-\Omega\right)\\
&-\frac{15}{32}e^2\sin^2i\sin^2\e\cos\left( 2\left(\lp+\omega\right)\right)\\
&-\frac{15}{32}e^2\sin^2i\sin^2\e\cos\left( 2\left(\lp-\omega\right)\right)\\
&-\frac{5}{64}e^2\left(1+\cos i\right)^2\left(1-\cos\e\right)^2\cos\left( 2\left(\lp+\Omega+\omega\right)\right)\\
&-\frac{5}{64}e^2\left(1-\cos i\right)^2\left(1-\cos\e\right)^2\cos\left( 2\left(\lp+\Omega-\omega\right)\right)\\
&-\frac{5}{64}e^2\left(1-\cos i\right)^2\left(1+\cos\e\right)^2\cos\left( 2\left(\lp-\Omega+\omega\right)\right)\\
&-\frac{5}{64}e^2\left(1+\cos i\right)^2\left(1+\cos\e\right)^2\cos\left( 2\left(\lp-\Omega-\omega\right)\right)\\
&-\frac{5}{16}e^2\sin i\left(1+\cos i\right)\sin\e\left(1-\cos\e\right)\\
&\cos\left( 2\left(\lp\right)+\Omega+2\omega\right)\\
&+\frac{5}{16}e^2\sin i\left(1-\cos i\right)\sin\e\left(1-\cos\e\right)\\
&\cos\left( 2\left(\lp\right)+\Omega-2\omega\right)\\
&-\frac{5}{16}e^2\sin i\left(1-\cos i\right)\sin\e\left(1+\cos\e\right)\\
&\cos\left( 2\left(\lp\right)-\Omega+2\omega\right)\\
&+\frac{5}{16}e^2\sin i\left(1+\cos i\right)\sin\e\left(1+\cos\e\right)\\
&\left.\cos\left( 2\left(\lp\right)-\Omega-2\omega\right)\right].
\end{split}
\end{equation}
From this Hamiltonian, we can identify 14 different possible combinations of angles between the orbital motion of the planet and the secular motion of the satellite.
The corresponding possible resonances are then 
\begin{eqnarray}
\nu_e \equiv \npr-\dot{\Omega}-\dot{\omega} & = & 0 \nonumber\\
\npr+\dot{\Omega}+\dot{\omega} & = & 0 \nonumber\\
\npr-\dot{\Omega} & = & 0 \nonumber\\
\nu_1 \equiv \npr+\dot{\Omega} & = & 0 \nonumber\\
\npr-\dot{\omega} & = & 0 \nonumber\\
\npr+\dot{\omega} & = & 0 \nonumber\\
\npr+\dot{\Omega}-\dot{\omega} & = & 0 \nonumber\\
\npr-\dot{\Omega}+\dot{\omega} & = & 0 \nonumber\\
2\npr-\dot{\Omega} & = & 0 \nonumber\\
\nu_2 \equiv 2\npr+\dot{\Omega} & = & 0 \nonumber\\
2\npr+\dot{\Omega}+2\dot{\omega} & = & 0 \nonumber\\
2\npr+\dot{\Omega}-2\dot{\omega} & = & 0 \nonumber\\
2\npr-\dot{\Omega}+2\dot{\omega} & = & 0 \nonumber \\
\nu_i \equiv 2\npr-\dot{\Omega}-2\dot{\omega} & = & 0 \ ,
\label{res_list}
\end{eqnarray}
where $\npr = \np - \dot \psi$ is the mean motion of the planet corrected by the precession frequency of the spin axis.

The argument of the pericenter, $\omega$, is not an inertial angle, as it is defined with respect to the line of nodes between the orbit of the satellite and the equator of the planet (Fig.~\ref{fig:orbit}), but it can be related with the inertial longitudes $\omega = \varpi - \Omega$.
The first resonance in the list (\ref{res_list}) thus becomes $\nu_e = \npr-\dot{\varpi} = 0$, and occurs when the pericenter precession frequency of the satellite, $\dot \varpi$, is equal to $\npr$.
This corresponds to the well-known evection resonance.
When capture in this resonance occurs, we can observe a significant increase in the eccentricity of the satellite \citep[e.g.,][]{touma1998,frouard2010}.
However, the amplitude of the evection resonance is proportional to $e^2$, and so this resonance cannot occur for satellites in nearly circular orbits, which is the final outcome of tidal evolution.
The resonance $\npr-\dot{\omega} = 0$ has been studied in the case of Phobos by \cite{yoder1982}, who showed that its crossing could lead to a sudden increase in the eccentricity of Phobos of maximal amplitude $0.016$.
We also note that in expression (\ref{eq:hamdev}) other secular resonances not involving the mean motion $\np$ are also possible, such as $\dot \omega = 0$ (Kozai resonance), but we do not study them here.

\section{Eviction-like resonances}
\label{evicition}

The last resonance in the list (\ref{res_list}), $\nu_i = 2\npr-2\dot{\varpi}+\dot{\Omega} = 0$, corresponds to the "eviction" resonance described by \citet{touma1998}, who showed that it can increase the inclination of the satellite.
In principle, any resonance containing the longitude of the ascending node, $\Omega$, can excite the inclination, and hence be dubbed as "eviction-like" resonance.
However, as for the evection resonance, the amplitudes of most of these resonances are also proportional to the square of the eccentricity (Eq.\,(\ref{eq:hamdev})), and so they are not present for nearly circular satellite orbits.
This incidentally includes the "original" eviction resonance $\nu_i$ reported by \citet{touma1998}.

In the Hamiltonian (\ref{eq:hamdev}), only four resonances, $\npr\pm\dot{\Omega} = 0$ and $2\npr\pm\dot{\Omega} = 0$, do not depend on the pericenter precession frequency, and thus do not have a null amplitude for an eccentricity equal to zero.
As a consequence, they are able to modify the inclination of a satellite even if the eccentricity is nearly zero, as currently observed for all tidally evolved satellites in the solar system.
Nonetheless, since $n_1>0$ and $\dot \Omega $ is expected to be negative for a prograde orbit,
only the resonances $\nu_1 = \npr+\dot{\Omega} = 0$ and $\nu_2 = 2\npr+\dot{\Omega} = 0$ can occur \citep{yokoyama2002}.
Therefore, in this section, we focus our study on these two resonances.

In general, we have $|\dot \psi |\ll \np$, thus, for simplicity, we assume $\dot{\psi}=0$ only in this section, such that we can use the Delaunay canonical variables.
It is possible to obtain these variables for a moving equator, but we needed to modify the Hamiltonian \citep[e.g.,][]{goldreich1965, kinoshita1993}.

\subsection{Resonance $\nu_1$}

We denote $\nu_1$ the resonance associated to the relation $\npr+\dot{\Omega} = 0$.
Among the resonant terms of the Hamiltonian (\ref{eq:hamdev}), we keep only the term associated with $\nu_1$, and we get for the resonant Hamiltonian
\begin{equation}
\begin{split}
\mathcal{H}_{\nu_1}=&\frac{\CR}{2a^3 \left(1-e^2\right)^{3/2}}\left(1-3\cos^2 i\right) + \frac{3\G m_\sun a^2}{2a_\orbit^3}\left(-\frac{e^2}{4}\right. \\
& +\frac{1}{4}\fe\left(\sin^2i+\sin^2\e-\frac{3}{2}\sin^2i\sin^2\e\right)\\
& \left.-\frac{1}{16}\fe\sin^2 i\left(1-\cos \e\right)^2\cos\left( 2\left(\lp+\Omega\right)\right)\right).
\end{split}
\label{eq:hamres}
\end{equation}
Using the Delaunay variables for the satellite $(L,G,H,l,g,h)$, where $L=\sqrt{\G m_\planet  a}$, $G=L\sqrt{1-e^2}$, $H=G\cos i$, $l=M$, $g=\omega$, and $h=\Omega$, we obtain
\begin{equation}
\begin{split}
&\mathcal{H}_{\nu_1}=\frac{\CR\left(\G m_\planet \right)^{3/2}}{2a^{3/2}G^3}\left(1-3\frac{H^2}{G^2}\right) + \frac{3\G m_\sun a^2}{2a_\orbit^3}\left(-\frac{L^2-G^2}{4L^2}\right. \\
& +\frac{1}{4}\left(1+\frac{3}{2}\frac{L^2-G^2}{L^2}\right)\left(\sin^2\e+\left(1-\frac{H^2}{G^2}\right)\left(1-\frac{3}{2}\sin^2\e\right)\right)\\
& -\frac{1}{16}\left(1+\frac{3}{2}\frac{L^2-G^2}{L^2}\right)\left(1-\frac{H^2}{G^2}\right)\\
& \left.\left(1-\cos \e\right)^2\cos\left( 2\left(\lp+h\right)\right)\right).
\end{split}
\end{equation}
The Hamiltonian $\mathcal{H}_{\nu_1}$ does not depend on the angles $l$ and $g$, and so the associated action variables $L$ and $G$ are constant.
As a result, the semi-major axis and the eccentricity are also constant, that is, the $\nu_1$ resonance does not modify the eccentricity.
By removing the constant terms and the terms which do not depend on $H$ or $h$, the Hamiltonian becomes
\begin{equation}
\mathcal{H}_{\nu_1}=-\mathcal{A}H^2-\mathcal{B}\left(G^2-H^2\right)\cos\left( 2\left(\lp+h\right)\right),
\end{equation}
with
\begin{equation}
\label{eq:coefA}
\mathcal{A}=\frac{3\G m_\sun a^2}{8G^2a_\orbit^3}\fe\left(1-\frac{3}{2}\sin^2\e\right)+\frac{3\CR}{2G^2a^3 \left(1-e^2\right)^{3/2}},
\end{equation}
\begin{equation}
\mathcal{B}=\frac{3\G m_\sun a^2}{32G^2a_\orbit^3}\fe\left(1-\cos \e\right)^2.
\end{equation}

As in \citet{touma1998} and \citet{yokoyama2002}, we use a generating function to perform a canonical change of variables, $F_1=(\lp+h)X$.
We obtain the canonical variables $(X,x)$ with $X=H$ and $x=\lp+h$. As we suppose $\dot \psi=0$ in this section, we have $\partial (\lp)/ \partial t=\np$, and the Hamiltonian becomes
\begin{equation}
\mathcal{H}_{\nu_1}=-\mathcal{A}X^2-\mathcal{B}\left(G^2-X^2\right)\cos 2x + \np X \ .
\label{Ham_canonique}
\end{equation}
The equations of motion are then given by
\begin{equation}
\dot{X}=-\frac{\partial \mathcal{H}_{\nu_1}}{\partial x}=-2\mathcal{B}\left(G^2-X^2\right)\sin 2x  \ ,
\end{equation}
\begin{equation}
\dot{x}=\frac{\partial \mathcal{H}_{\nu_1}}{\partial X}=-2\mathcal{A}X+2\mathcal{B}X\cos 2x + \np \ .
\end{equation}

\begin{figure}
\centering
\includegraphics[width=\columnwidth]{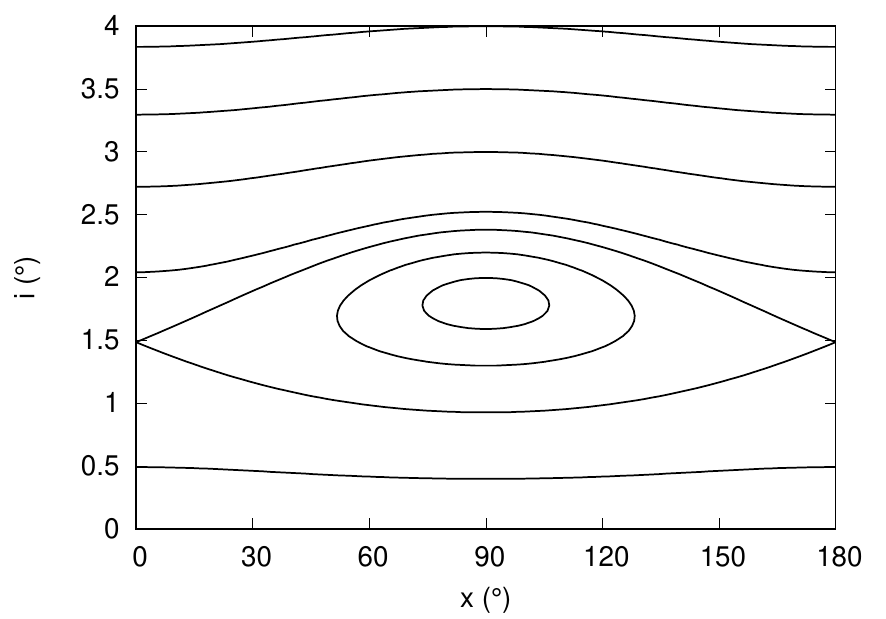}
\caption{Level curves of the Hamiltonian (\ref{Ham_canonique}) for Phobos (Table~\ref{Table_obsdata}), for a semi-major axis $a=2.61686\,R$, and assuming an obliquity $\e=90^\circ$ for Mars.}
\label{Alib_fig}
\end{figure}

\begin{figure}
\centering
\includegraphics[width=\columnwidth]{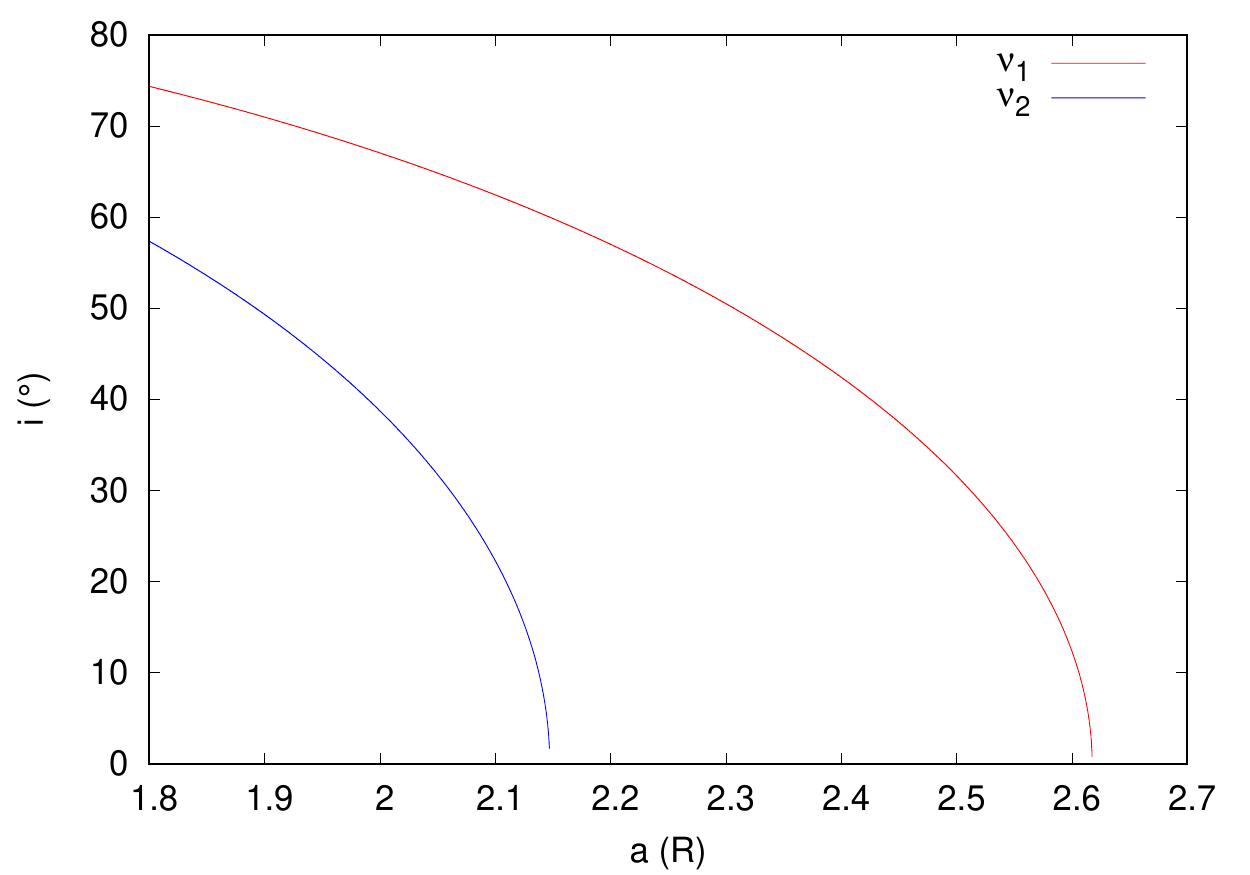}
\caption{Evolution of the inclination of Phobos for the stable fixed points with respect to its semi-major axis for the $\nu_1$ resonance (red curve) computed with Eq. (\ref{eq:n1stab}), and for the $\nu_2$ resonance (blue curve) computed with Eq. (\ref{eq:znu2}) for a Martian obliquity of $\e=25\degree$.
The Roche limit for Phobos is $a/R \approx 1.6$.}
\label{fig:equilibre}
\end{figure}

This Hamiltonian possesses four fixed points, obtained when $(\dot X, \dot x) = (0,0)$: the point $(X= \np/(2(\mathcal{A}+\mathcal{B})),x=\pi/2)$ and the point $(X=\np/(2(\mathcal{A}+\mathcal{B})),x=3\pi/2)$ are stable, while the points $(X=\np/(2(\mathcal{A}-\mathcal{B})),x=0)$ and $(X=\np/(2(\mathcal{A}-\mathcal{B})),x=\pi)$ are unstable.
In Fig.~\ref{Alib_fig} we show the level curves of the Hamiltonian (\ref{Ham_canonique}) for Phobos, assuming an obliquity $\e=90^\circ$ for Mars (Table~\ref{Table_obsdata}), where we can clearly identify the fixed points as well as the separatrix that encircles the resonant area.
If the satellite is at a stable fixed point, its inclination verifies 
\begin{equation}
\label{eq:n1stab}
\cos i= \frac{\np}{2G\left(\mathcal{A}+\mathcal{B}\right)} \ ,
\end{equation}
while for an unstable fixed point, the inclination is given by
\begin{equation}
\cos i= \frac{\np}{2G\left(\mathcal{A}-\mathcal{B}\right)} \ .
\label{eq:n1unstab}
\end{equation}
Therefore, the stable points are present only if $\np\leq2G(\mathcal{A}+\mathcal{B})$, and the unstable ones if $\np\leq2G(\mathcal{A}-\mathcal{B})$.
The two equilibrium points are usually very close to each other.
Indeed, near the $\nu_1$ resonance, the perturbation due to the flattening of the planet is in general much larger than the stellar perturbation, and so
\begin{equation}
\mathcal{A}+\mathcal{B}\approx\mathcal{A}-\mathcal{B}\approx\frac{3\CR}{2G^2a^3(1-e^2)^{3/2}} \ .
\end{equation}
In Fig.~\ref{fig:equilibre}, we show the stable equilibria for Phobos computed with the values from Table~\ref{Table_obsdata}, assuming an obliquity of Mars similar to the present one, $\e=25^\circ$.
We observe that there is a critical value for the semi-major axis of the satellite (obtained with $\cos i = 1$),
\begin{equation}
a_{\nu_1}\approx\left(\frac{3\CR}{\sqrt{\G m_\planet}\np(1-e^2)^{2}}\right)^{2/7},
\label{eq:ac}
\end{equation}
above which the $\nu_1$ resonance is not present.
In the case of Phobos, we have $a_{\nu_1}/R \approx 2.617$.
This value is very close to the surface of Mars, but still larger than the Roche limit for a solid body with the density of Phobos, $a/R \approx 1.61$ \citep[e.g.,][]{Chandrasekhar_1987}.

\begin{table}
\caption{Values of the present parameters of Phobos, Mars, and the Sun used in this work.
For some parameters, we give an approximate value.}
\centering
\begin{tabular}{|c|c|c|}
\hline
parameter & unit & value \\
\hline
$\G m_0$ & $\AUd$ & $2.959122 \times 10^{-4}$ \\
$\G m_1$ & $\AUd$ & $9.5495351\times 10^{-11}$ \\
$\G m\,^{(a)}$ & $\mathrm{km}^{3}/\mathrm{s}^{2}$ & $7.092 \times 10^{-4}$ \\
$a$ & $\mathrm{km}$ & $9375.$ \\
$e$ & $-$ & $0.015$ \\
$i$ & $\degree$ & $1.$ \\
$T_1$ & $\mathrm{day}$ & $686.98$ \\
$n_1$ & rad/day & $2\pi/T_1$ \\
$a_1$ & $\mathrm{au}$ & $(\G (m_0+m_1)/n_1^2)^{1/3}$ \\
$\e$ & $\degree$ & $25.$ \\
$\dot{\psi}\,^{(b)}$ & $\arc$ & $-7.6083$ \\
$J_2\,^{(b)}$ & $-$ & $1.95661 \times 10^{-3}$ \\
$R\,^{(b)}$ & $\mathrm{km}$ & $3396.$ \\
$k_2\,^{(c)}$ & $-$ & $0.183$ \\
$Q\,^{(c)}$ & $-$ & $99.5$\\
\hline
\end{tabular}      
\tablefoot{$J_2 = (C-\frac12(A+B))/(m_1R^2)$; $^{(a)}$ \citet{jacobson2010}; $^{(b)}$ \citet{konopliv2016}; $^{(c)}$ \citet{jacobson2014}.}
\label{Table_obsdata}
\end{table}

\subsection{Resonance $\nu_2$}

We denote $\nu_2$ the resonance associated to the relation $2\npr+\dot{\Omega} = 0$.
Among the resonant terms of the Hamiltonian (\ref{eq:hamdev}), we keep only the term associated with $\nu_2$, and we get for the resonant Hamiltonian
\begin{equation}
\begin{split}
\label{Ham_nu2}
&\mathcal{H}_{\nu_2}=\frac{\CR}{2a^3 \left(1-e^2\right)^{3/2}}\left(1-3\cos^2 i\right) + \frac{3\G m_\sun a^2}{2a_\orbit^3}\left(-\frac{e^2}{4}\right. \\
& +\frac{1}{4}\fe\left(\sin^2i+\sin^2\e-\frac{3}{2}\sin^2i\sin^2\e\right)\\
& \left.+\frac{1}{8}\fe\sin\left(2i\right)\sin\e\left(1-\cos\e\right)\cos\left( 2\left(\lp\right)+\Omega\right)\right).
\end{split}
\end{equation}
As for the $\nu_1$ resonance, the Hamiltonian does not depend on the angle $g$ and the eccentricity remains constant.
The Hamiltonian then becomes
\begin{equation}
\mathcal{H}_{\nu_2}=-\mathcal{A}H^2+\mathcal{C}H\sqrt{G^2-H^2}\cos\left( 2\left(\lp\right)+h\right),
\end{equation}
where $\mathcal{A}$ is given by expression (\ref{eq:coefA}), and
\begin{equation}
\mathcal{C}=\frac{3\G m_\sun a^2}{8G^2a_\orbit^3}\fe\sin \e\left(1-\cos \e\right).
\end{equation}
We now consider the generating function $F_2=(2(\lp)+h)Y$.
The canonical variables $(Y,y)$ verify $Y=H$ and $y=2(\lp)+h$, and the resonant Hamiltonian becomes
\begin{equation}
\mathcal{H}_{\nu_2}=-\mathcal{A}Y^2+\mathcal{C}Y\sqrt{G^2-Y^2}\cos y+2\np Y \ .
\end{equation}
The equations of motion are then given by
\begin{equation}
\dot{Y}=-\frac{\partial \mathcal{H}_{\nu_2}}{\partial y}=\mathcal{C}Y\sqrt{G^2-Y^2}\sin y,
\end{equation}
\begin{equation}\label{eq:ydotnu2}
\dot{y}=\frac{\partial \mathcal{H}_{\nu_2}}{\partial Y}=-2\mathcal{A}Y+\mathcal{C}\frac{G^2-2Y^2}{\sqrt{G^2-Y^2}}\cos y +2 \np \ .
\end{equation}
The fixed points are then located at $y=0$ or $y=\pi$.
In order to obtain the $Y$  values for the fixed points, we must solve the following quartic equation
\begin{equation}
z^4-2bcz^3-\left(1-b^2\right)z^2+2bcz+\frac{1-c^2}{4}-b^2=0,\label{eq:znu2}
\end{equation}
with $z=Y/G$, $c=\mathcal{A}/\sqrt{\mathcal{A}^2+\mathcal{C}^2}$, and $b=\np/(G\sqrt{\mathcal{A}^2+\mathcal{C}^2})$.
Contrary to the $\nu_1$ resonance, it is not possible to find a simple analytical expression for the fixed points, although we can find the roots numerically.
In Fig.~\ref{fig:equilibre}, we show the stable fixed point located at $y=\pi$ for Phobos with the values from Table~\ref{Table_obsdata}, and assuming an obliquity of Mars similar to the present one, $\e=25^\circ$.
For a satellite close to the $\nu_2$ resonance, the perturbations due to the flattening of the planet are in general larger than the stellar perturbations, and so $\mathcal{A}\approx3\CR/(2G^2a^3(1-e^2)^{3/2})\gg\mathcal{C}$.
It is then possible to obtain an approximate value for the inclination of the fixed points.
By neglecting the term proportional to $\mathcal{C}$ in Eq. (\ref{eq:ydotnu2}), we obtain
\begin{equation}
-2\mathcal{A}Y+2 \np =0 \ ,
\end{equation}
and the inclination is given by
\begin{equation}
\cos i \approx \frac{\np}{G\mathcal{A}} \ .
\label{eq_nu2}
\end{equation}
Therefore, as for the $\nu_1$ resonance, there is a critical value $a_{\nu_2}$ of the semi-major axis above which the $\nu_2$ resonance is not present.
An approximate value of $a_{\nu_2}$ is given by (obtained with $\cos i = 1$)
\begin{equation} \label{acprime}
a_{\nu_2} \approx\left(\frac{3\CR}{2\sqrt{\G m_\planet}\np(1-e^2)^{2}}\right)^{2/7}\approx \frac{a_{\nu_1}}{2^{2/7}} \approx 0.82 \, a_{\nu_1} \ ,
\end{equation}
where $a_{\nu_1}$ is the critical value for the $\nu_1$ resonance (Eq.\,(\ref{eq:ac})).
In the case of Phobos, we have $a_{\nu_2}/R \approx 2.147$.

\subsection{Capture probabilities}
\label{sec_pcap}

Due to tidal effects, the semi-major axis of the satellites usually evolves with time.
When the semi-major axis is increasing, the eviction-like resonances are crossed if we initially have $a<a_\nu$.
As a result, capture is not possible, because the resonances are no longer present for $a>a_\nu$ (Fig.~\ref{fig:equilibre}).
As noted by \cite{touma1998}, the crossing of the resonance can nevertheless lead to a sudden variation in the inclination of the satellite due to the conservation of the area in the phase plane (Fig.~\ref{Alib_fig}).

On the other hand, when the semi-major axis is decreasing, the eviction-like resonances are encountered if we initially have $a>a_\nu$.
As a result, capture in resonance may occur.
In that case, as the semi-major axis is decreasing, the inclination increases (Fig.~\ref{fig:equilibre}).
We then conclude that, for a satellite with an orbit close to the equator, capture in eviction-like resonances is only possible if the semi-major axis is decreasing.

It is possible to compute analytically the capture probability in the $\nu_1$ resonance, provided that the evolution of the semi-major axis is adiabatic.
To be considered as adiabatic, the evolution of the semi-major axis must be much slower than the conservative inclination variations, that is, tidal dissipation must be weak.
In general, in the phase space of a given resonance we can distinguish three different regions delimited by the separatrix of the resonance (Fig.~\ref{Alib_fig}): a libration zone, with area $\Alib$, and two circulation zones, one above the libration zone, with area $A_{circ}$, and another below the libration zone, with area $A_{circ}'$.
The capture probability is then obtained by the modification of the phase space with time, that is, by the change in the areas encircled by the separatrix \citep{yoder1979,henrard1982,henrard1993}
\begin{equation}
P_\mathrm{cap}=\dAlib / \dAtot \ ,\label{eq:p_C}
\end{equation}
where $\Atot$ corresponds to the sum of the libration area with the circulation area increasing with time.
The ordinate of the stable fixed point inside the libration area is (Eq.\,(\ref{eq:n1stab}))
\begin{equation}
X = \frac{\np}{2(\mathcal{A}+\mathcal{B})} \approx \frac{\np G^2a^3(1-e^2)^{3/2}}{3\CR} \propto a^4 \ ,
\end{equation}
and so the ordinate of the fixed point decreases when the semi-major decreases, which means that the circulation area whose surface increases with time is $A_{circ}$, and then $\Atot=\Alib+A_{circ}$.

As the phase space of the $\nu_1$ resonance is $\pi-$periodic (Eq.\,(\ref{Ham_canonique})), we restrict the computation of the capture probabilities to $x \in [0:\pi]$.
We perform the following canonical change of variables $(X,x)\rightarrow(z,x)$ with $z=X/G = \cos i$.
With these variables, the Hamiltonian (\ref{Ham_canonique}) becomes
\begin{equation}
\mathcal{H}'_{\nu_1}=-\mathcal{A}Gz^2-\mathcal{B}G\left(1-z^2\right)\cos 2x +\np z.
\end{equation}
To compute the libration area, we need to know the equation of the separatrix.
The unstable fixed point (\ref{eq:n1unstab}) is on the separatrix, whose energy is
\begin{equation}
\mathcal{H}'_{S}=\left(\mathcal{D}-\mathcal{B}\right)G \ , \quad \mathrm{with} \quad
\mathcal{D}= \frac{\np^2}{4G^2\left(\mathcal{A}-\mathcal{B}\right)} \ .
\end{equation}
The equation of the separatrix is then
\begin{equation}
-\mathcal{A}Gz^2-\mathcal{B}G\left(1-z^2\right)\cos 2x +\np z=\left(\mathcal{D}-\mathcal{B}\right)G \ ,
\end{equation}
and so, the equations of the superior separatrix, $z_{+}$, and the inferior one, $z_{-}$, are
\begin{equation}
\begin{split}
z_\pm = & \frac{\np}{2G\left(\mathcal{A}-\mathcal{B}\cos 2x\right)}\\
& \pm\frac{\sqrt{2\mathcal{B}\sin^2x\left(\mathcal{A}-\mathcal{B}\cos 2x-\mathcal{D}\right)}}{\mathcal{A}-\mathcal{B}\cos 2x}
\end{split} \ .\label{eq:separatrix}
\end{equation}

The libration area encircled by the separatrix is
\begin{equation}
\begin{split}
\Alib & = \int_{0}^{\pi}z_+\,dx-\int_{0}^{\pi}z_-\,dx \\
& =2\sqrt{2\mathcal{B}}\int_{0}^{\pi} \frac{\sqrt{\sin^2x\left(\mathcal{A}-\mathcal{B}\cos 2x-\mathcal{D}\right)}}{\mathcal{A}-\mathcal{B}\cos 2x}\,dx \\
& =2\sigma \sqrt{\frac{2}{\tau}}\int_{0}^{\pi}\frac{\sin\left(\frac{x}{2}\right)\sqrt{1-\tau\cos x}}{1-\sigma\cos x}\,dx  \ ,
\end{split}
\label{alib_int}
\end{equation}
with
\begin{equation}
\sigma=\frac{\mathcal{B}}{\mathcal{A}} 
\quad \mathrm{and} \quad
\tau=\frac{\sigma}{1-\mathcal{D}\mathcal{A}^{-1}} \ .
\end{equation}
In order that the separatrix and the libration area exist, we must have $\np< 2G(\mathcal{A}-\mathcal{B})< 2G(\mathcal{A}+\mathcal{B})$.
Therefore, $\sigma$ and $\tau$ verify $0<\sigma<\tau<1$.
We obtain 
\begin{equation}
\begin{split}
&\int_{0}^{\pi}\frac{\sin\left(\frac{x}{2}\right)\sqrt{1-\tau\cos x}}{1-\sigma\cos x}\,dx=  \frac{1}{\sigma\sqrt{2}}\left(\sqrt{\tau}\arccos\left(1-\frac{4\tau}{1+\tau}\right)\right. \\
& \left.-\sqrt{\frac{\tau-\sigma}{1+\sigma}}\arccos\left(1-\frac{4\left(\tau-\sigma\right)}{\left(1-\sigma\right)\left(1+\tau\right)}\right)\right) \ ,
\end{split}
\end{equation}
and finally
\begin{equation}
\begin{split}
\Alib= & 2\left(\arccos\left(1-\frac{4\tau}{1+\tau}\right)\right. \\
& \left. -\sqrt{\frac{\tau-\sigma}{\tau\left(1+\sigma\right)}}\arccos\left(1-\frac{4\left(\tau-\sigma\right)}{\left(1-\sigma\right)\left(1+\tau\right)}\right)\right).
\end{split}
\label{eq:Alib}
\end{equation}

For the total area we get
\begin{equation}
\begin{split}
\Atot &=\int_{0}^{\pi}\,dx-\int_{0}^{\pi}z_-\,dx \\
& = \pi -\int_{0}^{\pi}\frac{\np}{2G\left(\mathcal{A}-\mathcal{B}\cos 2x\right)}dx+\frac{\Alib}{2} \\
& =\pi\left(1-\frac{\np}{2G\sqrt{\mathcal{A}^2-\mathcal{B}^2}}\right)+\frac{\Alib}{2} \ .
\end{split}
\label{eq:Atot}
\end{equation}

If we consider only the tidal effects on the semi-major axis, in the expressions of $\Alib$ and $\Atot$ only the semi-major axis depends on the time.
Then, from expression (\ref{eq:p_C}) the capture probability is given by 
\begin{equation}
P_\mathrm{cap}=\frac{\partial \Alib}{\partial a} \Big/ \frac{\partial \Atot}{\partial a} \ ,
\label{eq:p_C2}
\end{equation}
which does not depend on the dissipation law chosen for the semi-major axis.
From the expressions of $\Alib$ (Eq.\,(\ref{eq:Alib})), and of $\Atot$ (Eq.\,(\ref{eq:Atot})), we can compute the derivatives with respect to the semi-major axis, and finally the capture probability.

\begin{figure}
\centering
\includegraphics[width=\columnwidth]{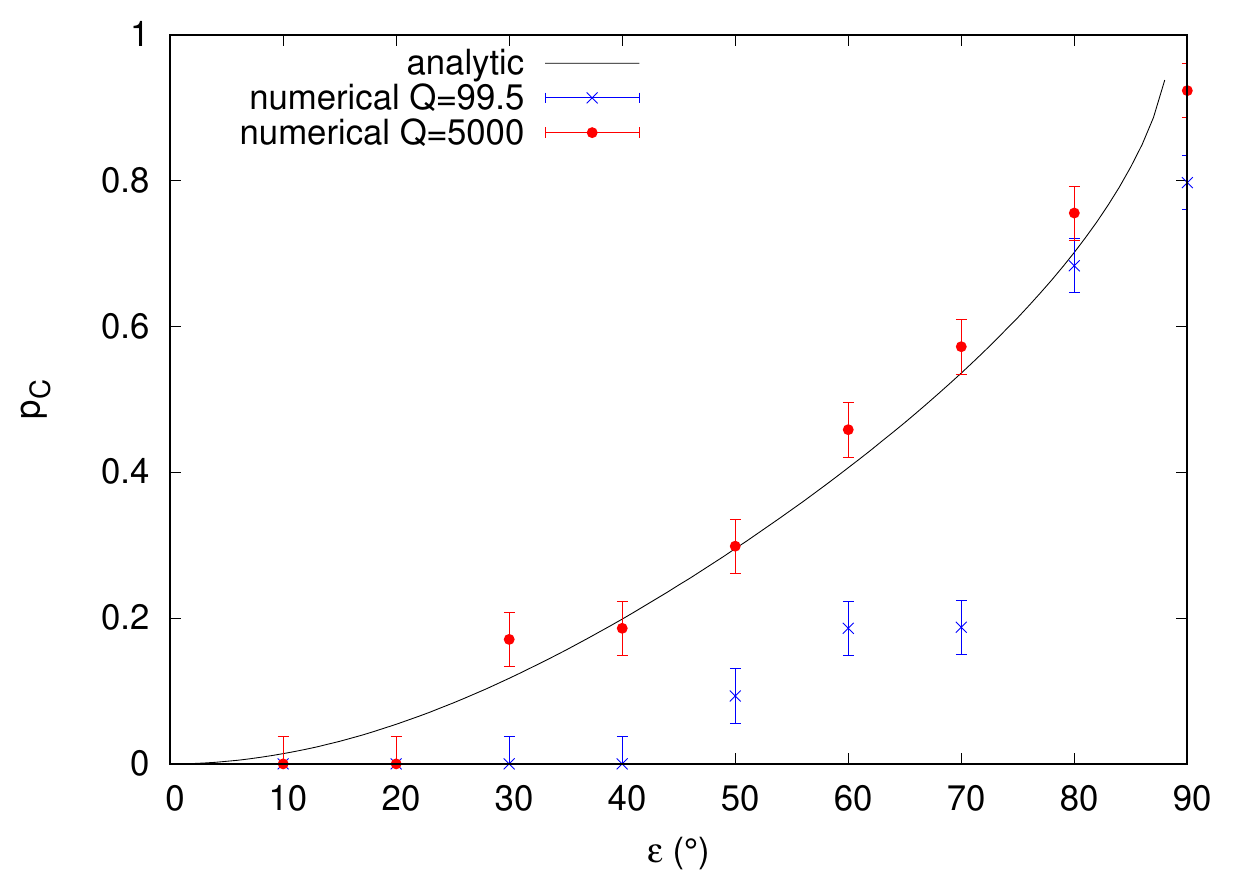}
\caption{Capture probability of Phobos in the $\nu_1$ resonance with respect to the obliquity of Mars.
The black curve corresponds to the capture probability computed with Eq. (\ref{eq:p_C2}).
The blue crosses and the red dots correspond to the capture probability determined for each value of the obliquity with $N=720$ numerical integrations of the resonant Hamiltonian (Eq. (\ref{eq:hamres})),  respectively with $Q=99.5$ and $Q=5000$ for the effective specific tidal dissipation.
The errorbars are given by $\pm1/\sqrt{N}$.}
\label{pcap_figure}
\end{figure}

In Fig.~\ref{pcap_figure}, we show the capture probability in the $\nu_1$ resonance for Phobos as a function of the obliquity of Mars (with initial values taken from Table~\ref{Table_obsdata}).
The solid line gives the theoretical estimation obtained with expression (\ref{eq:p_C2}), using a critical semi-major axis obtained with expression (\ref{eq:n1stab}).
In order to obtain a numerical estimation of the capture probability with different values of the tidal dissipation (for more details see Eq.\,(\ref{sma_eq}) and Sect.~\ref{num_pcap_sec}), we perform numerical integrations of the resonant Hamiltonian (\ref{eq:hamres}).
We modify this Hamiltonian following \cite{goldreich1965} and \cite{kinoshita1993} to consider the motion of the Martian equator, which we model here with a constant obliquity and a uniform precession.

For the present tidal dissipation of Mars ($Q=99.5$), there are some differences between the numerical results and the analytic computation.
However, for weaker tidal dissipation ($Q=5000$), which corresponds to a slower evolution of the semi-major axis of Phobos (adiabatic evolution), we observe that there is a good agreement.
We hence conclude that the evolution of Phobos cannot be considered as adiabatic.

\begin{figure}
\centering
\includegraphics[width=\columnwidth]{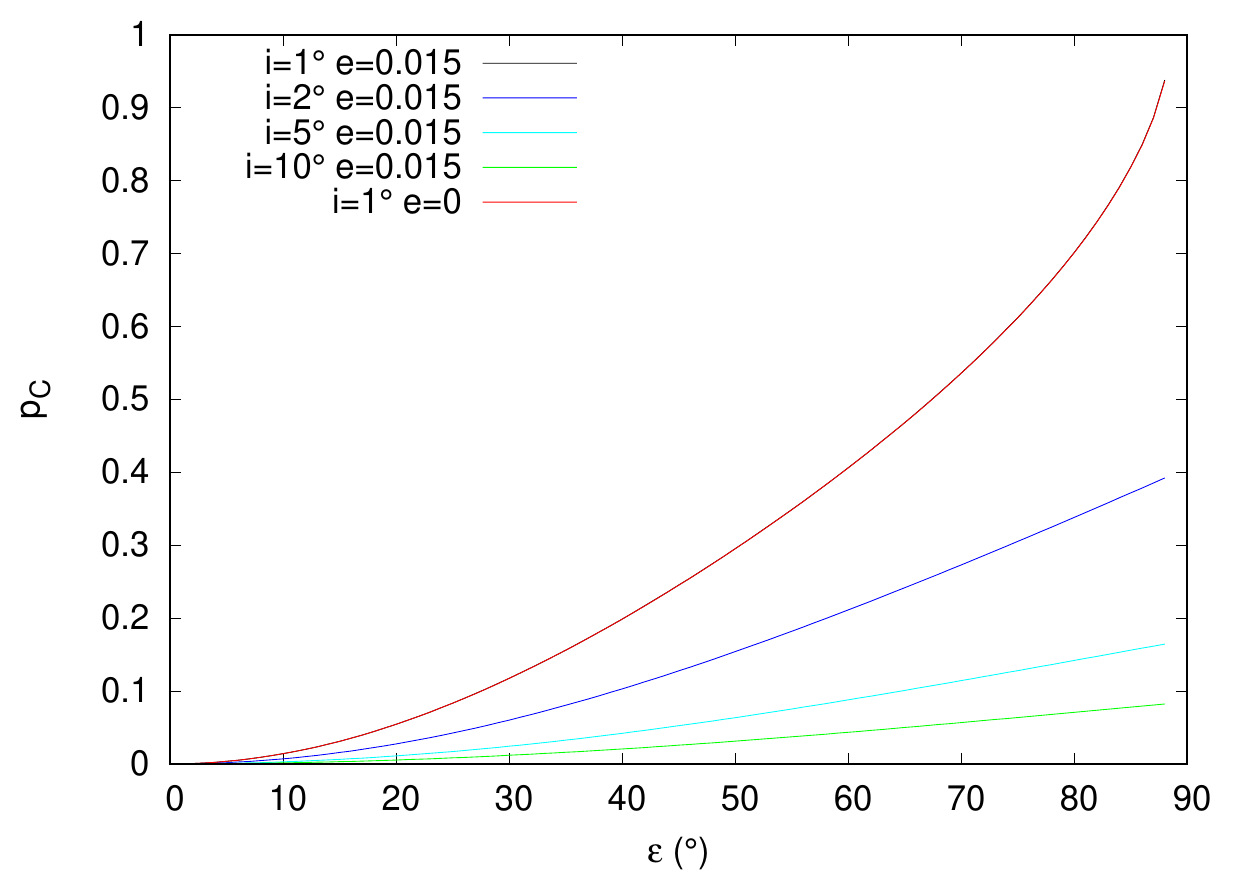}
\caption{Capture probability of Phobos in the $\nu_1$ resonance with respect to the obliquity of Mars for different orbital parameters of Phobos.
The black line corresponds to the current parameters.}
\label{fig:proba_analytic}
\end{figure}

In Fig.~\ref{fig:proba_analytic}, we show the theoretical capture probability (Eq.\,(\ref{eq:p_C2})) also as function of the obliquity of Mars, but for different inclinations and eccentricities of Phobos.
As expected, we observe that the capture probability increases with the obliquity, because for small obliquity, $\e$, the amplitude of the $\nu_1$ resonance is proportional to $\e^4$  (Eq. (\ref{eq:hamres})).
On the other hand, we observe that the capture probability strongly decreases with the inclination of Phobos.
The red curve, obtained with $e=0$, is very close to the black curve, obtained with $e=0.015$, that is, the capture probability is not much influenced by a small variation in eccentricity.
This was also expected, since the amplitude of the $\nu_1$ resonance is proportional to $1+3e^2/2$  (Eq. (\ref{eq:hamres})).

The Hamiltonian of the $\nu_1$ resonance is similar to the one of the second order ivection resonance studied by \cite{xufabrycky2019} in the case of a binary star system.
The resonances called ivection by \cite{xufabrycky2019} correspond to a resonance between the node precession frequencies of planets orbiting around a star with the mean motion of the binary.
At the lowest order in the mutual inclination of planets, \cite{xufabrycky2019} obtained that the Hamiltonian of the second order ivection resonance can be written in the same dimensionless form as the $j+2:j$ orbital resonance described by \cite{borderies1984}.
In the limit of small inclination, it is then also possible to use the analytical expression for the capture probability given by \cite{borderies1984}.

The capture probabilities in the $\nu_2$ resonance can be obtained with a similar method as for the $\nu_1$ resonance. 
However, unfortunately it is not possible to find a simple analytical expression for the corresponding $\Alib$ integral (Eq.\,(\ref{alib_int})).
At the lowest order in inclination of the satellite, the $\nu_2$ resonance and the first order ivection resonance of \cite{xufabrycky2019} can be written in the same dimensionless form as the $j+1:j$ orbital resonance described by \cite{borderies1984}.
In the limit of small inclination, the analytical expression of the capture probability given by \cite{borderies1984} can then also be used for the $\nu_2$ resonance.

\section{Interaction between resonances}
\label{interaction}

In previous section, we have studied the dynamics of an isolated resonance.
However, in the full Hamiltonian (\ref{eq:hamdev}) there is a large number of resonant terms that may interact.
This is particularly true, when two resonant islands are close to each other, as it is the case of the $\nu_1$ resonance and the evection resonance $\nu_e$ \citep{yokoyama2002}.
Indeed, in general we have $\dot \varpi \approx - \dot \Omega$, which is the case for a satellite with a weakly inclined orbit with respect to the equator of its planet.
Thus, for nonzero eccentricities, the libration width of these two resonances may overlap and trigger chaotic motion \citep{Chirikov_1979}.
In order to describe this interaction, we need to add to the previous $\nu_1$ resonant Hamiltonian (\ref{eq:hamres}) the term corresponding to the evection resonance $\nu_e$.
The interaction Hamiltonian is then
\begin{equation}
\begin{split}
&\mathcal{H}_{int}=\frac{\CR}{2a^3 \left(1-e^2\right)^{3/2}}\left(1-3\cos^2 i\right) + \frac{3\G m_\sun a^2}{2a_\orbit^3}\left(-\frac{e^2}{4}\right. \\
& +\frac{1}{4}\fe\left(\sin^2i+\sin^2\e-\frac{3}{2}\sin^2i\sin^2\e\right)\\
& -\frac{1}{16}\fe\sin^2 i\left(1-\cos \e\right)^2\cos\left( 2\left(\lp+\Omega\right)\right)\\
& \left.-\frac{5}{64}e^2\left(1+\cos i\right)^2\left(1+\cos\e\right)^2\cos\left( 2\left(\lp-\varpi\right)\right)\right).
\end{split}
\label{eq:hamint}
\end{equation}

\subsection{Frequency analysis method}

The interaction Hamiltonian has two degrees of freedom, and it is thus not integrable.
This kind of problem is often studied using surface sections, but here we propose a different method: we use stability maps based on frequency analysis \citep{laskar1988,laskar1990,laskar1992,laskar1993,laskar2003}.
This method decomposes a discrete temporal function in a quasi-periodic approximation and can estimate its stability.
We consider a grid of initial conditions and integrate the equations of motion over a time $T$.
Then, we perform a frequency analysis\footnote{We analyze the quantity $k'_x+ik'_y$, where $\vb{k}'=(k_x,k_y,k_z)$ corresponds to the coordinates of the normal to the satellite's orbit expressed in the referential of the equator of the planet. In the quasi-periodic approximation, the precession frequency corresponds to the main frequency with the largest amplitude.}
over the time intervals $[0:T/2]$ and $[T/2:T]$, and determine the precession frequency of the ascending node of the satellite in each interval, $f_1$ and $f_2$, respectively.
The stability is measured by 
\begin{equation}
\sigma \equiv \left\lvert1 - \frac{f_2}{f_1}\right\rvert \ ,
\end{equation}
which estimates the chaotic diffusion of the precession frequency \citep{dumas1993,laskar1993}.
The larger $\sigma$ is, the more unstable the orbital motion of the satellite is.
For stable motion we have $\sigma \sim 0$, while $\sigma \ll 1$ if the motion is weakly perturbed, and $\sigma \sim 1$ when the motion is chaotic.
It is difficult to know precisely what is the value of $\sigma$ for which the motion is stable or unstable, but a threshold of stability $\Dlim$ can be estimated such that most of the trajectories with $\sigma < \Dlim$ are stable \citep{couetdic2010}.
For $\sigma < \Dlim$, we still would like to distinguish the circulation trajectories from those that are in resonance.
For that purpose, in those cases we compute a second quantity
\begin{equation}
\theta \equiv \left\lvert1+\frac{f_1}{\npr}\right\rvert \ ,
\label{sigma_resonant}
\end{equation}
which measures the relative difference between the precession frequency and the resonance frequency $-\npr$.
We assume that a trajectory is in resonance when $ \theta < \theta_r$, where the boundary $\theta_r$ is chosen such that it is able to correctly identify the libration area in the case of the resonant Hamiltonian (\ref{eq:hamres}).
In the stability maps, we attribute a color scheme to the different $\sigma$ going from blue (stable) to red (unstable), while for the resonant trajectories we apply a black filter.

\subsection{Stability maps}

As for surface sections, the frequency analysis is a numerical method, and thus we need to attribute values to the parameters of the system.
We adopt here the values for Phobos and Mars (Table~\ref{Table_obsdata}) near the $\nu_1$ resonance, for which $a_{\nu_1}/R \approx 2.617$.
This resonance is more interesting than the $\nu_2$ resonance, since it is close to the evection resonance $\nu_e$.
We also fix the obliquity of Mars at $\e=90\degree$, because the resonance width of $\nu_1$ is larger for this obliquity (Eq.\,(\ref{eq:hamres})) for $\e\in[0\degree:90\degree]$.
We build a 2D mesh of initial conditions where the inclination of the satellite varies from $0.04\degree$ to $4\degree$ with a step size of $0.04\degree$, and the canonical variable $x=\lp+\Omega$ varies from $0\degree$ to $180\degree$ with a step size of $2\degree$, and we integrate the equations of motion over the time interval $T=80$~kyr.

\begin{figure*}
\centering
\begin{tabular}{ccc}
resonant Hamiltonian $\mathcal{H}_{\nu_1}$ (Eq.\,(\ref{eq:hamres})) \quad \quad &
interaction Hamiltonian $\mathcal{H}_{int}$ (Eq.\,(\ref{eq:hamint})) & \quad
secular Hamiltonian $\mathcal{H}_s$ (Eq.\,(\ref{eq:ham1}))
\end{tabular} \\
\includegraphics[width=6.cm,height=4.cm]{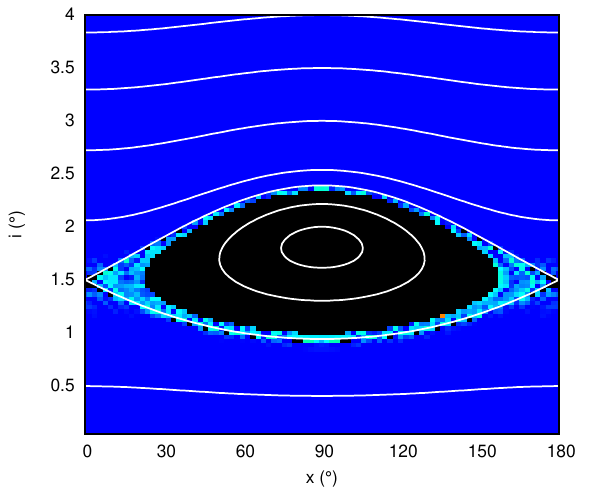}
\includegraphics[width=6.cm,height=4.cm]{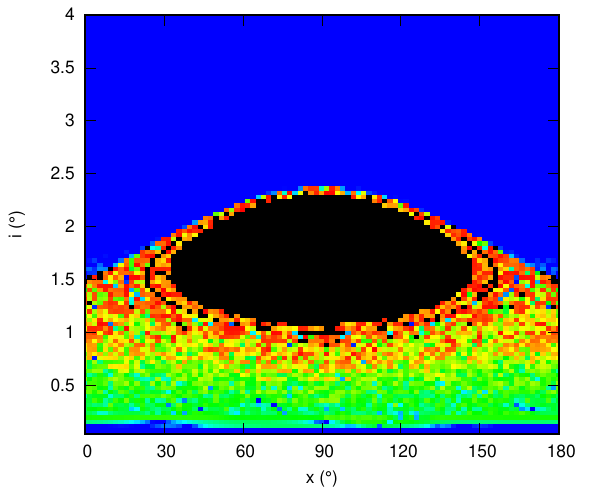} 
\includegraphics[width=6.cm,height=4.cm]{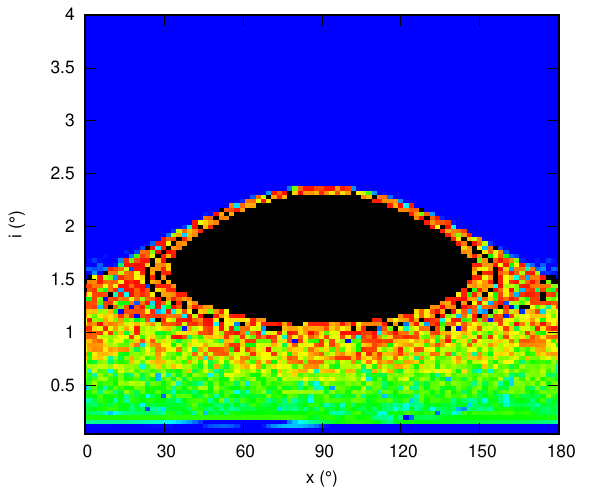} \\
\includegraphics[width=6.cm,height=4.cm]{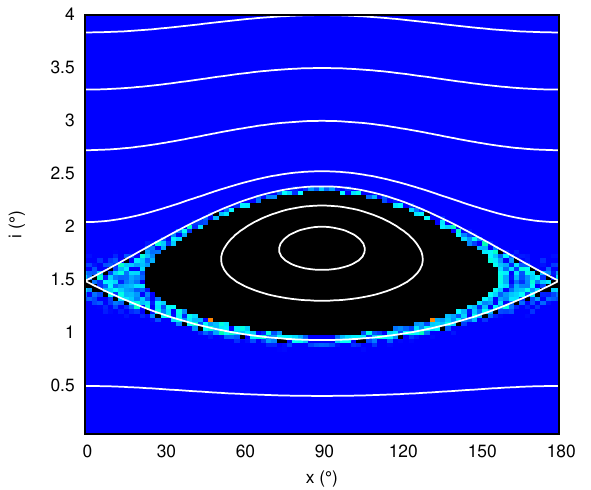}
\includegraphics[width=6.cm,height=4.cm]{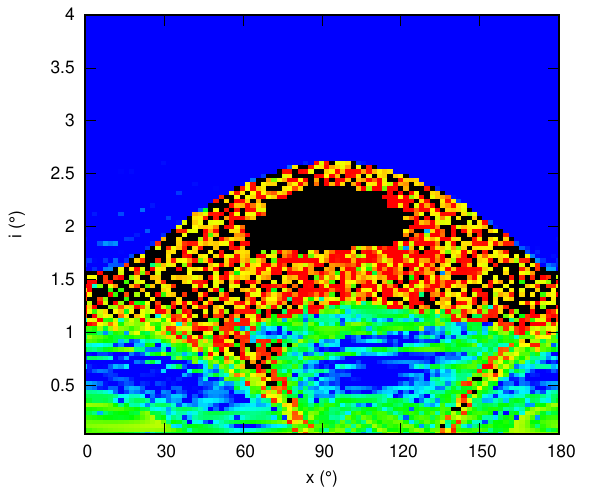} 
\includegraphics[width=6.cm,height=4.cm]{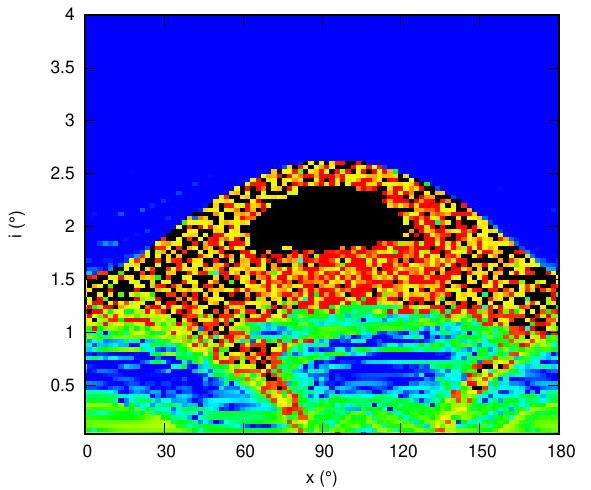} \\
\includegraphics[width=6.cm,height=4.cm]{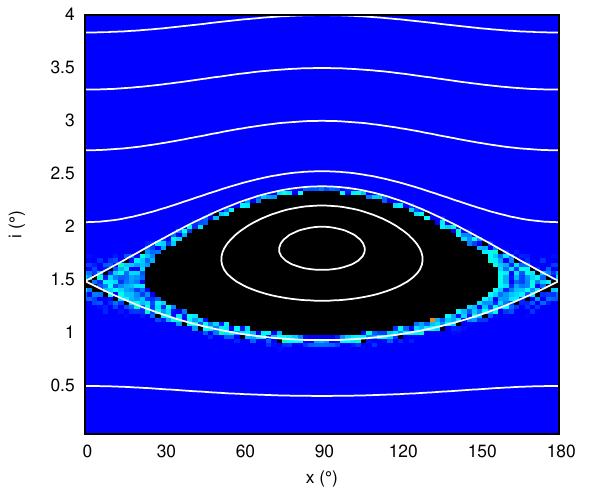}
\includegraphics[width=6.cm,height=4.cm]{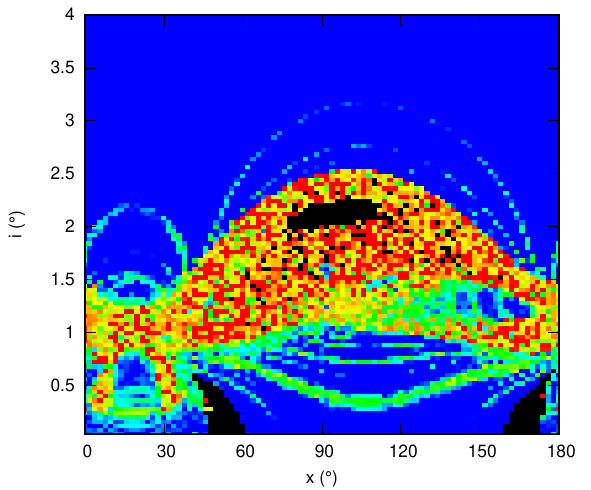} 
\includegraphics[width=6.cm,height=4.cm]{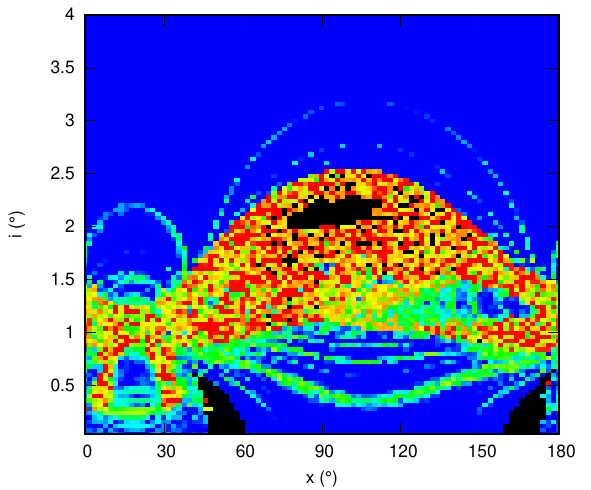} \\
\includegraphics[width=6.cm,height=4.cm]{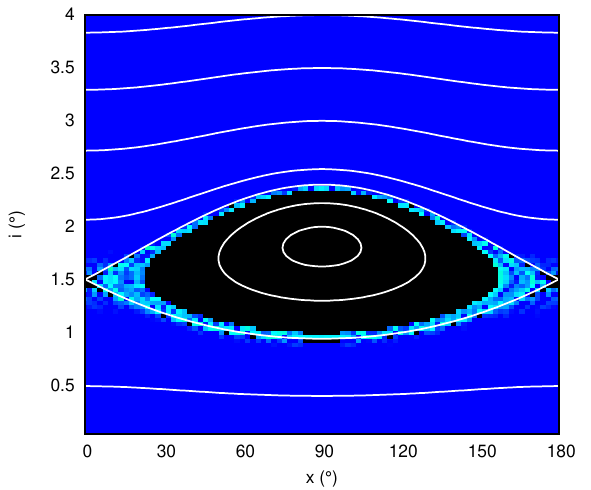}
\includegraphics[width=6.cm,height=4.cm]{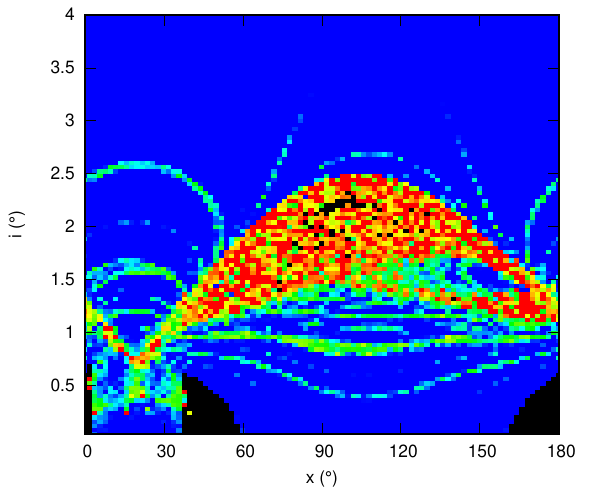} 
\includegraphics[width=6.cm,height=4.cm]{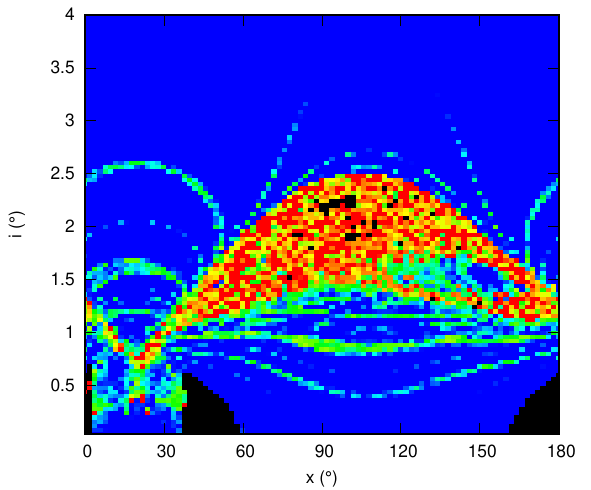} \\
\includegraphics[width=\columnwidth]{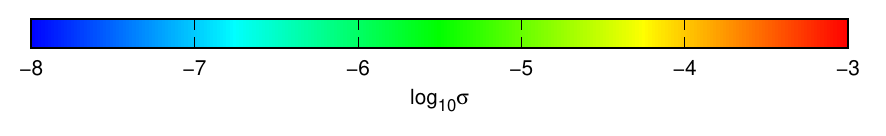}
\caption{Stability maps for Phobos (Table~\ref{Table_obsdata}) using an obliquity for Mars $\e=90\degree$. From top to bottom, the initial eccentricity is $e = 10^{-5}$, $e=0.015$, $e=0.03$, $e=0.045$. Each column corresponds to the integration using different Hamiltonians: resonant Hamiltonian (left), interaction Hamiltonian (middle) and the full secular Hamiltonian (right). 
The color scale indicates the value of the quantity $\log_{10} \sigma$. The blue and green areas correspond to stable circulating trajectories, while the yellow, orange, and red ones correspond to chaotic motions. The black areas correspond to stable resonant regions, for which $ \log_{10} \theta < -3.9$. \label{fig:mape0}}
\end{figure*}

In Fig.~\ref{fig:mape0}, we show the stability maps obtained for different initial values of the eccentricity of Phobos $e=10^{-5}$, 0.015, 0.030, and 0.045, since the amplitude of the evection resonance increases with the eccentricity (Eq.\,(\ref{eq:hamint})).
For each value of the initial eccentricity, we numerically integrate the equations of motion obtained for the full secular Hamiltonian $\H_s$ (Eq. (\ref{eq:ham1})), for the resonant Hamiltonian $\mathcal{H}_{\nu_1}$ (Eq.\,(\ref{eq:hamres})), and for the interaction Hamiltonian $\mathcal{H}_{int}$ (Eq.\,(\ref{eq:hamint})).
For the three cases, we consider that the Martian obliquity is constant, and that the precession of the Martian equator is uniform.
To consider the motion of the Martian equator, we modify the resonant Hamiltonian $\mathcal{H}_{\nu_1}$ and the interaction Hamiltonian $\mathcal{H}_{int}$ following \cite{goldreich1965} and \cite{kinoshita1993}.
That is, we obtain three different maps for each eccentricity value\footnote{As the location of the stable point of the $\nu_1$ resonance depends on the eccentricity (Eq.\,(\ref{eq:n1stab})), and also on the considered Hamiltonian, we need to slightly modify the value of the semi-major axis in order that the location of the stable point is located about at the same place for each eccentricity and Hamiltonian.}: the left-hand map corresponds to the integrable resonant Hamiltonian for the $\nu_1$ resonance (Eq. (\ref{eq:hamres})), the middle map corresponds to the sum of this resonant Hamiltonian with the term due to the evection resonance (Eq. (\ref{eq:hamint})), and the right-hand map illustrates the true dynamics of the system obtained with the full secular Hamiltonian (Eq. (\ref{eq:ham1})).
The resonant Hamiltonian (Eq. (\ref{eq:hamres})) gives identical trajectories for the initial conditions $x$ and $x+\pi$, that is why we only consider $x \in [0^\circ, 180^\circ]$.
This symmetry is broken when we introduce the perturbations from the other resonant terms, but the changes are almost imperceptible. 

We attribute a color scheme to the different values of $\sigma$, and estimate for the stability threshold\footnote{Following \citet{couetdic2010}, we estimate the diffusion $\sigma'$ for a shorter interval $[0:T']$, where $T'=T/100$. In general, for a stable trajectory $\sigma<\sigma'$, because the accuracy of the frequency analysis increases with the size of the interval, while $\sigma>\sigma'$ for an unstable trajectory, as the chaotic diffusion increases with time. We then construct the histogram of the trajectories verifying $\sigma<\sigma'$ with a step of $0.5$. The values of $\sigma$ and $\sigma'$ are averaged over the closest neighbors of the grid of initial conditions to decrease the effects of chaotic trajectories which have small values of the diffusion. From this histogram, we determine with a linear interpolation the value $\log_{10} \Dlim=-4.4$ below which $99\%$ of the trajectories verify $\sigma<\sigma'$.}, $\log_{10} \Dlim=-4.4$.
Therefore, the blue and green areas correspond to stable circulating trajectories, while the yellow, orange, and red ones correspond to chaotic motions.
In the panels corresponding to the resonant Hamiltonian (left-hand maps in Fig.~\ref{fig:mape0}) we additionally show the level curves of constant energy obtained with expression (\ref{eq:hamres}), as this problem is integrable (see also Fig.~\ref{Alib_fig}).
We observe there is a good agreement between these curves and the stability map obtained with frequency analysis.
These panels thus allow us to calibrate the transition between the libration and circulation areas.
We find that trajectories inside the separatrix are bounded by $ \log_{10} \theta_r = -3.9$ (Eq.\,(\ref{sigma_resonant})).
We then apply a black filter for stable trajectories with $\theta < \theta_r$ that correspond to resonant regions\footnote{For the maps with $e=0.03$ and $e=0.045$ in Fig. \ref{fig:mape0}, some black regions appear around $i=0\degree$ and $i=1\degree$ that do not correspond to the $\nu_1$ resonance. They correspond to stable circulation regions that incidentally also have $\theta < \theta_r$.}.

In Fig.~\ref{fig:mape0}, we observe that as we increase the eccentricity, the dynamics changes considerably between the resonant case and the full problem.
However, a first striking evidence is that the evection resonance $\nu_e$ is the major responsible for this modification, since the middle maps obtained only with the interaction Hamiltonian (Eq. (\ref{eq:hamint})) are very similar to the right-hand maps obtained with the full secular Hamiltonian (Eq. (\ref{eq:ham1})).
This means that the dynamics of Phobos around the $\nu_1$ resonance is essentially governed by the interaction between $\nu_1$ and $\nu_e$.
More detailed studies of this problem, thus only need to take the two-degree of freedom Hamiltonian (\ref{eq:hamint}) into account.

As expected, for a nearly zero eccentricity, $e=10^{-5}$ (first row in Fig.~\ref{fig:mape0}), all maps are similar to the one obtained with the resonant Hamiltonian (\ref{eq:hamres}).
Indeed, for such a small eccentricity, the amplitude of the evection resonance $\nu_e$ is nearly zero, and the dynamics is dominated by the $\nu_1$ resonance.
Nonetheless, in the case of the full Hamiltonian we observe already some diffusion of the orbits in the region around the libration area, because the eccentricity is not exactly zero.
For the present eccentricity of Phobos, $e = 0.015$ (second row in Fig.~\ref{fig:mape0}), the $\nu_1$ resonance still presents a large libration area in the full Hamiltonian case, but is now surrounded by a large chaotic region.
Its chaotic nature is notably characterized by a high fickleness of the value of the diffusion index, $\sigma$.
For an eccentricity of $e=0.03$ (third row in Fig.~\ref{fig:mape0}), the libration area corresponding to the $\nu_1$ resonance still exists, but it is considerably smaller and deformed, while for $e=0.045$ (last row in Fig.~\ref{fig:mape0}) the $\nu_1$ resonance almost disappeared in the middle of a chaotic region.
Finally, we also observe for the full secular Hamiltonian and the resonant Hamiltonian that as the eccentricity increases, the center of the libration area is slightly shifted toward increasing values of $x> 90^\circ$.

In Fig.~\ref{fig:mape0}, we confirm that the interaction between the resonances $\nu_1$ and $\nu_e$ increases with the eccentricity.
For the present eccentricity of Phobos (or smaller), this interaction is not very important.
As a result, in the future Phobos may survive in the $\nu_1$ resonance.
However, as the eccentricity increases, we observe that the chaotic region around $\nu_1$ progressively replaces the libration area.
It is thus unlikely for a satellite with eccentricities larger than $0.05$ to be captured in this resonance.

\section{Application to Phobos}
\label{phobos}

In order to maximize the impact of the eviction-like resonances on the inclination of a satellite, the semi-major axis must decrease.
Tidal interactions between the satellite and the planet can account for this evolution, provided that the orbital period of the satellite is shorter than the rotational period of the planet, which is the case of Phobos, the largest satellite of Mars \citep[e.g.,][]{szeto1983}.
Among the main satellites of solar system planets, only Triton, the largest satellite of Neptune, is also spiraling into the planet, but because it is on a retrograde orbit \citep[e.g.,][]{Correia_2009}.
Therefore, in this section we apply our model to study Phobos.

At present, the semi-major axis of Phobos is $a/R = 2.761$, and the eccentricity is $e=0.01511$ \citep{jacobson2014}.
Among the fourteen resonances listed in Eq. (\ref{res_list}), seven resonances can act on a prograde satellite such as Phobos.
During its past evolution, Phobos already encountered four of these resonances, but their effect on its orbit is presumably to have been negligible \citep{yokoyama2002}, since
the amplitudes of these resonances are proportional to the square of the eccentricity.
In the future, Phobos will encounter the $\nu_e$ and the $\nu_1$ resonances, which occur almost simultaneously around $a_{\nu_1}/R = 2.617$, and later the $\nu_2$ resonance, which is located at $a_{\nu_2}/R = 2.147$.
For the $\nu_1$ and the $\nu_2$ resonances, the amplitudes are not null for eccentricities close to zero, and so they can modify the future inclination of Phobos.

In Sect. \ref{sec_pcap}, we have obtained the capture probability with an analytical computation.
However, expression (\ref{eq:p_C2}) is only valid for the $\nu_1$ resonance and if the evolution is adiabatic.
Moreover, it does not take the interaction between the $\nu_1$ resonance and the evection resonance $\nu_e$ into account, that we have seen in Sect. \ref{interaction}.
It is then necessary to perform numerical simulations to capture the correct dynamics while crossing the $\nu_1$ and $\nu_2$ resonances.
Here we investigate the capture probabilities in these two resonances, and determine which parameters can influence this capture.
In the case where capture is not possible, we also study the effects due to the crossing of these resonances on the orbital evolution of Phobos.

\subsection{Numerical model}

We numerically integrate the set of equations (\ref{eq:num1}) to obtain the secular orbital evolution of Phobos forced by the orbital motion of Mars and its equator.
We consider that the orbit of Mars in the ecliptic is circular and that its equator has a uniform precession and a constant obliquity.

Tides raised by Phobos on Mars, modify all the orbital parameters of the satellite with time.
However, for simplicity we consider only the evolution of the semi-major axis.
We also do not consider tides raised by Mars on Phobos, whose effect is very small, since the rotation of Phobos is currently synchronous and on a near circular orbit.
In order to describe the evolution of the semi-major axis, we adopt a constant$-Q$ tidal model \citep[e.g.,][]{szeto1983},
\begin{equation}
\dvt{a}=-3\sqrt{\frac{\G}{m_1}}\frac{mR^5}{a^{11/2}} \frac{k_2}{Q},
\label{sma_eq}
\end{equation}
where $m$ is the mass of Phobos, $R$ is the radius of Mars, $k_2$ is the second Love number, and $Q$ is the effective specific tidal dissipation.
From the observation of the satellites of Mars, it is possible to put constraints on the coefficients $k_2$ and $Q$.
We adopt the values obtained by \cite{jacobson2014} with this method: $k_2=0.183$ and $Q=99.5$ (Table~\ref{Table_obsdata}).

\subsection{Numerical setup}
\label{num_pcap_sec}

The capture probability in eviction-like resonances depends on the inclination of the satellite and on the obliquity of the planet (Sect. \ref{sec_pcap}).
However, it can also depend on the eccentricity, because of the interaction with neighbor resonances (Sect. \ref{interaction}).
The current eccentricity of Phobos is about $0.015$, but it is still being damped to smaller values.
It is then likely that it will be smaller when Phobos approaches the resonances.
Then, in order to observe the impact of the eccentricity in the capture probability, we adopt different eccentricity values up to 0.015.

The obliquity of Mars is chaotic and cannot be precisely computed beyond about 10~Myr \citep{touma1993,laskarrobutel1993,laskar2004}.
It experiences variations between $0\degree$ and $60\degree$ over 50~Myr, and can even reach values beyond $60\degree$ over higher periods of time.
For instance, over 100~Myr, the obliquity has a probability of $7.23\%$ to reach $60\degree$ and cannot have values beyond $70\degree$, but over 5~Gyr, the probabilities for the obliquity to reach $60\degree$, $70\degree$, and $80\degree$, are respectively $95.35\%$, $8.51\%$, and $0.015\%$ \citep{laskar2004}.
Due to the decrease in the semi-major axis, it is expected that Phobos will collide with Mars in about 40~Myr \citep{Efroimsky_Lainey_2007}.
It is then very unlikely that Mars has an obliquity higher than $60\degree$ when Phobos encounters the eviction-like resonances. 
We nevertheless investigate the probability of capture for all obliquity values up to $90\degree$, in order to get a global vision of this mechanism.

To numerically estimate the capture probability in resonance, for each initial value of the eccentricity and obliquity we perform $720$ simulations varying the initial longitude of the ascending node $\Omega$ from $0\degree$ to $359.5\degree$ with a step of $0.5\degree$, while fixing the remaining initial parameters.

The rotational and orbital motion of Mars is complex, but can be approximated with quasi-periodic solutions \citep[e.g.,][]{ward1979,laskar1990,laskarrobutel1993,laskar2004}.
In such solutions, the Martian obliquity is not constant, but the current dominant term corresponds to rapid oscillations, which can induce variations between $10.8\degree$ and $38.0\degree$ with a period of about 120~kyr \citep{ward1979}.
Therefore, its evolution time-scale is much longer than the precession motion of Phobos, and so it can be considered constant during the resonance crossing.
We consider a fixed orbit for Mars and a uniform precession of the Martian equator, that is, without taking the planetary perturbations into account that induce secular variations in periods much longer than the precession motion of Phobos.
We also consider a circular orbit for Mars to simplify the problem.

\subsection{Resonance $\nu_1$\label{sec:proba_n1}}

\subsubsection{Capture probabilities}

\begin{figure}[t]
\centering
\includegraphics[scale=0.7]{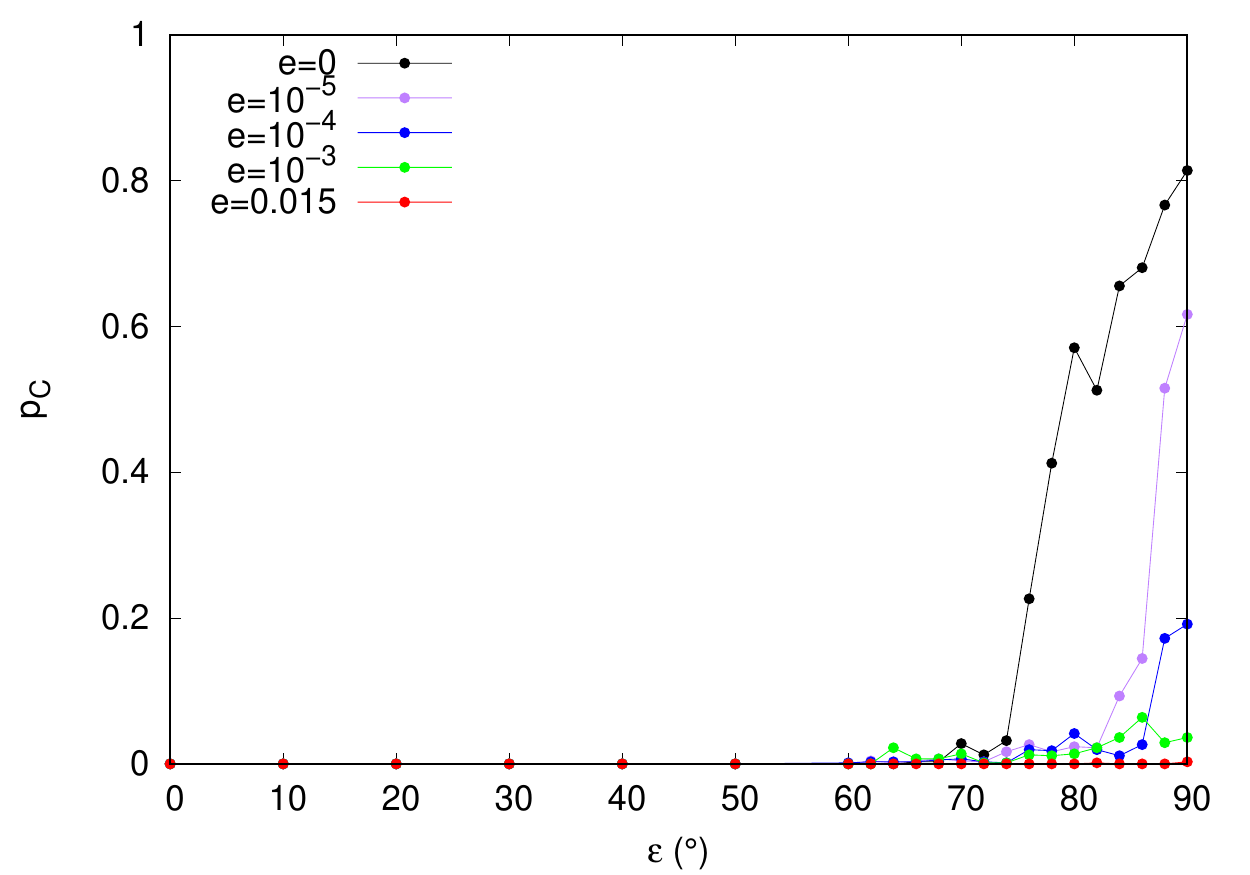}
\caption{Capture probabilities of Phobos in the $\nu_1$ resonance with respect to the obliquity of the equator of Mars for different initial eccentricities.\label{fig:probanum_n1}}
\end{figure}

We compute the capture probabilities in the $\nu_1$ resonance  using the present inclination of Phobos,  $i=1\degree$.
For the eccentricity, we adopt values $e=0$, $10^{-5}$, $10^{-4}$, $10^{-3}$ and $0.015$, while for the obliquity we adopt values from $0\degree$ to $60\degree$ with a step of $10\degree$, and from $60\degree$ to $90\degree$ with a step of $2\degree$, since the variations in the capture probability are more important when the obliquity is high (Sect. \ref{sec_pcap}).
We start our simulations for a semi-major axis slightly larger than $a_{\nu_1}$ (Eq.\,(\ref{eq:ac})) and integrate the equations of motion for 500~kyr.
The results are shown in Fig.~\ref{fig:probanum_n1}.

As expected, we observe that the capture probability in the $\nu_1$ resonance increases with the obliquity (Sect.~\ref{sec_pcap}).
For all the considered eccentricities, we did not observe any capture for an obliquity smaller than $60\degree$, and when $\e=60\degree$, a unique capture occurred (for $e=10^{-4}$).
The probability of capture for $e=0$ increases from $0.28\%$ for $\e=66\degree$ to $81.4\%$ for $\e=90\degree$.
For an initial eccentricity of $0.015$, which corresponds to the current value of the eccentricity, the probability of capture is zero except for  $\e=82\degree$ and $\e=90\degree$, where one and two captures have been observed, which correspond to capture probabilities of about $0.14\%$ and $0.28\%$, respectively.
These results are consistent with those observed by \cite{yokoyama2002}, but present some disagreement with the theoretical estimations done in Sect.~\ref{sec_pcap}.
One reason is because tides acting on Phobos semi-major axis are very strong (Eq.\,(\ref{sma_eq})), and hence the evolution is not adiabatic (see Fig.~\ref{pcap_figure}).
The other reason is the perturbations from the evection resonance (Sect.~\ref{interaction}).

\begin{figure}[t]
\centering
\includegraphics{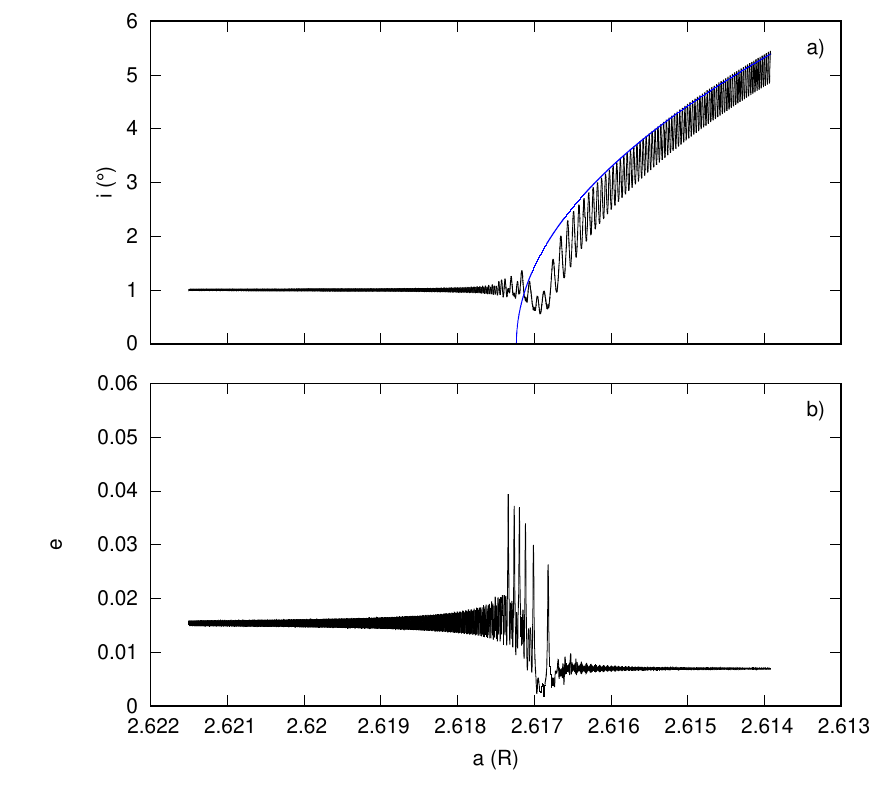}
\caption{Evolution of the inclination (a) and of the eccentricity (b) during the capture in the $\nu_1$ resonance for an obliquity of $\e=90\degree$ with the initial conditions $e=0.015$ and $\Omega=351.5\degree$.
The blue curve corresponds to the stable fixed point of the $\nu_1$ resonance computed with Eq. (\ref{eq:n1stab}).
\label{fig:inccap}}
\end{figure}

\begin{figure}[t]
\centering
\includegraphics{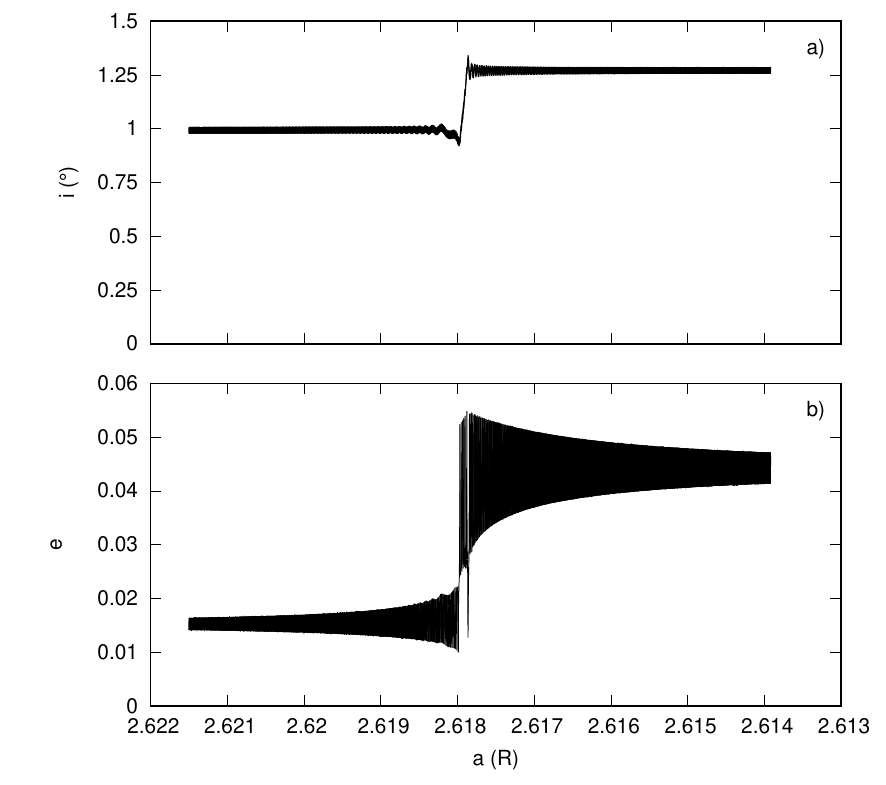}
\caption{Evolution of the inclination (a) and of the eccentricity (b) during the crossing of the $\nu_1$ resonance for an obliquity of $\e=60\degree$ with the initial conditions $e=0.015$, and $\Omega=231\degree$.
\label{fig:incnoncap_inc}}
\end{figure}

\begin{figure}[t]
\centering
\includegraphics{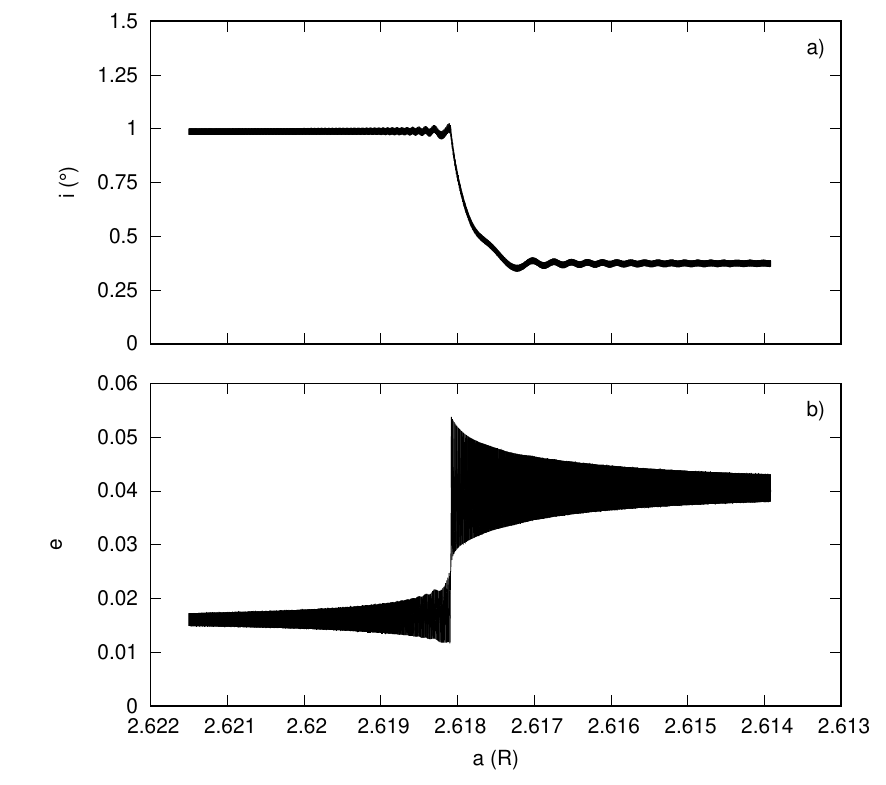}
\caption{Evolution of the inclination (a) and of the eccentricity (b) during the crossing of the $\nu_1$ resonance for an obliquity of $\e=60\degree$ with the initial conditions $e=0.015$,  and $\Omega=203\degree$.
\label{fig:incnoncap_dec}}
\end{figure}

According to expression (\ref{eq:hamres}), the influence of the eccentricity should be weak, because the resonance width slightly increases with the eccentricity.
However, in Fig.~\ref{fig:probanum_n1}, we note that the probability of capture sharply decreases when the eccentricity slightly increases.
For instance, for an obliquity of $90\degree$, the probability is $81.4\%$ for $e=0$, $61.7\%$ for $e=10^{-5}$, $19.2\%$ for $e=10^{-4}$, and $3.6\%$ for $e=10^{-3}$.
This is somehow surprising, because in Fig.~\ref{fig:mape0} we observe that for the present eccentricity ($e=0.015$) a large resonance island is still present.
However, as seen in previous sections, the $\nu_1$ resonance is very close to the evection resonance $\nu_e$.
As the semi-major axis of Phobos is decreasing, this resonance occurs almost simultaneously with the $\nu_1$ resonance.
Capture in the evection resonance is not possible, but crossing it induces a variation in the eccentricity of Phobos as we can see in Figs.~\ref{fig:inccap}, \ref{fig:incnoncap_inc}, and \ref{fig:incnoncap_dec}.

\begin{figure}[t]
\centering
\includegraphics{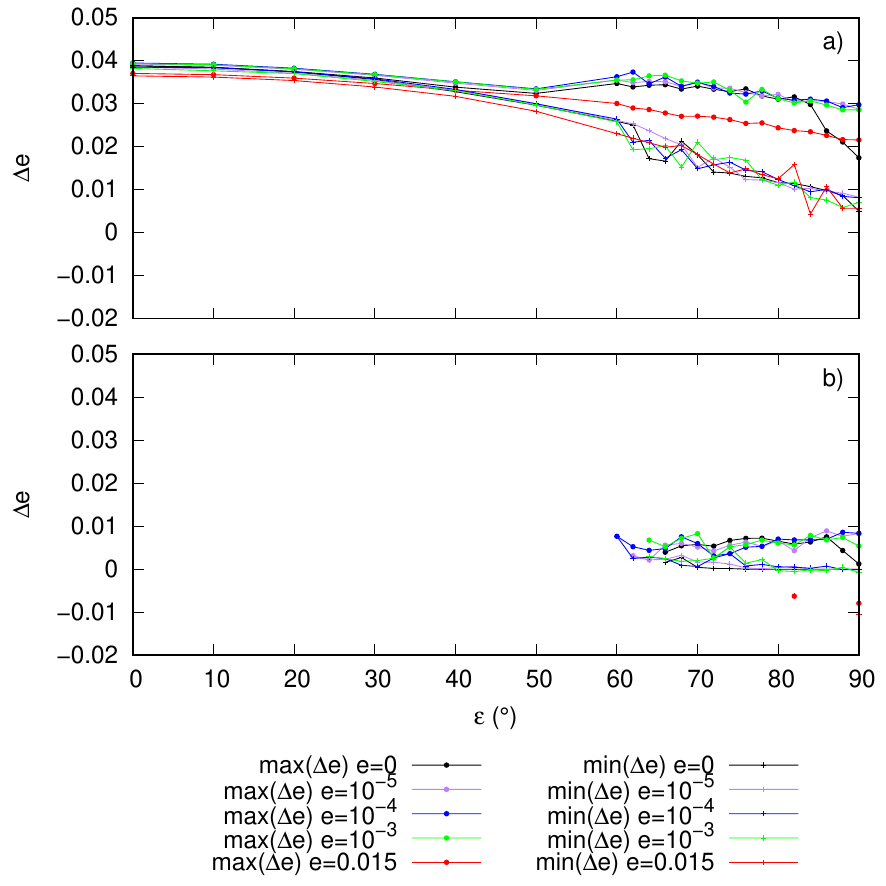}
\caption{Evolution of the maximum and minimum variations in the mean eccentricity $\Delta e$ owing to the encounter with the $\nu_1$ resonance in the cases where the capture in the resonance fails (a), and succeeds (b), with respect to the obliquity for different initial eccentricities.\label{fig:excnum_n1}}
\end{figure}

In Fig.~\ref{fig:excnum_n1}, we plot the maximum and minimum differences, $\Delta e$, between the mean eccentricity before and after the encounter with the $\nu_1$ resonance as a function of the obliquity, when the capture succeeds and fails.
If the capture fails, regardless of the initial value of the eccentricity before encountering the $\nu_1$ resonance, the final mean eccentricity always increases.
We note that during the crossing of the $\nu_1$ resonance the eccentricity can even take higher values before decreasing to a smaller final value (Figs. \ref{fig:incnoncap_inc}, \ref{fig:incnoncap_dec}).
For small obliquities, the eccentricity is always strongly excited by about $0.04$ (e.g., for $\e=0\degree$, we have  $0.0365 < \Delta e < 0.0395$), which is consistent with the results previously observed by \citet{yoder1982} and \citet{yokoyama2002}.
For obliquities larger than $60\degree$, we observe a smaller average variation around $0.02$, that can nevertheless still temporarily increase to about $0.04$ (e.g., for $\e=90\degree$, we have $0.005 < \Delta e < 0.0298$).
For these larger eccentricity values, the libration width of the $\nu_1$ resonance is much smaller and the capture probability is hence considerably reduced (see Fig.~\ref{fig:mape0}).

In Fig.~\ref{fig:excnum_n1}, we also show the statistics when capture succeeds\footnote{We do not have many examples for the initial eccentricity $0.015$ after capture, because the probability is very small.}.
In this case, we observe smaller variations in the eccentricity.
Therefore, the chances of capture in the $\nu_1$ resonance are weak because of the nearby evection resonance $\nu_e$.
This resonance leads to an increase in the eccentricity, which in turn gives rise to a chaotic region around the $\nu_1$ resonance that reduces its libration width (see Sect.~\ref{interaction}).

We conclude that capture in the $\nu_1$ resonance can only occur for very small eccentricity values and for an obliquity higher than $60\degree$.
These chances are on the low side, because as we have just seen, the eccentricity of Phobos will likely increase after crossing the evection resonance, and the obliquity of Mars is not expected to exceed $60\degree$ in the next 40~Myr.
It is then extremely unlikely that in the future Phobos gets caught in the $\nu_1$ resonance.
Nevertheless, if ever capture occurs, this resonance is stable and the inclination can increase.
Indeed, in Fig.~\ref{fig:inccap} we show an example of this capture and subsequent evolution for $e=0.015$ and $\e=90^\circ$.
We also confirm that, when capture occurs, the inclination follows the equilibrium point given by expression (\ref{eq:n1stab}).

\subsubsection{Inclination variation\label{sec:incnum}}

When capture in resonance does not occur, we still observe inclination variations during the resonance crossing.
As for the capture probability, the exact behavior of the inclination depends on the initial value of the longitude of the ascending node.
In Figs.~\ref{fig:incnoncap_inc} and~\ref{fig:incnoncap_dec}, we show two examples of such variations in the case of the $\nu_1$ resonance, which correspond to an increase and to a decrease in the initial inclination, respectively.
These variations modify the state of a satellite and can lead to misinterpretations of its origin.
We can profit from the simulations already performed to estimate the capture probabilities to put limits on the inclination variations.
Moreover, the $\nu_2$ resonance is encountered after the $\nu_1$ resonance, and so it is important to estimate the inclination of Phobos after crossing the $\nu_1$ resonance, because the capture probability depends on the inclination (Sect. \ref{sec_pcap}).

\begin{figure}[t]
\centering
\includegraphics[scale=0.7]{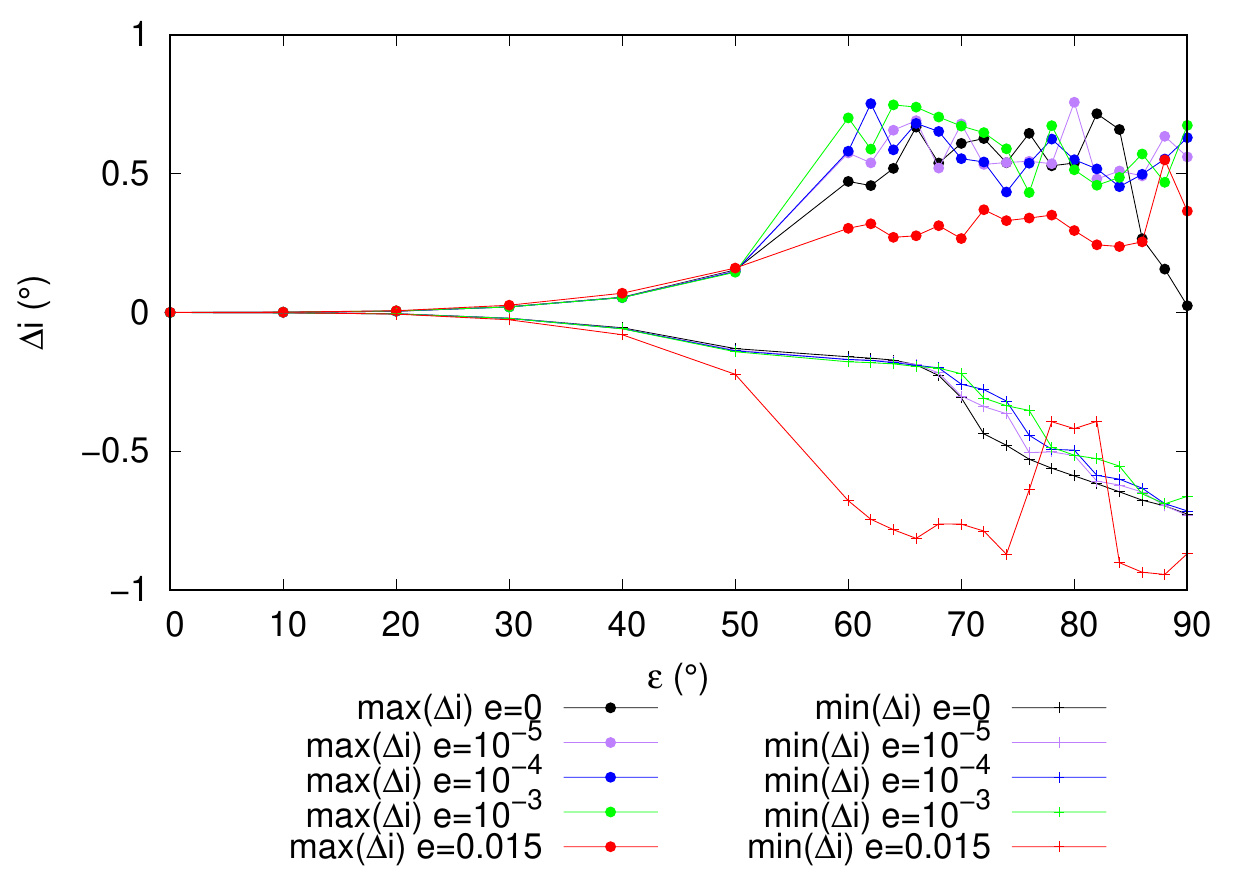}
\caption{Evolution of the maximum and minimum variations in the mean inclination $\Delta i$ owing to the encounter with the $\nu_1$ resonance in the case where the capture in the resonance fails, with respect to the obliquity for different initial eccentricities.\label{fig:incnum_n1}}
\end{figure}

In Fig.~\ref{fig:incnum_n1} we show the maximum and minimum differences between the mean inclination of Phobos before and after the crossing of the $\nu_1$ resonance.
As for the capture probability, we observe that these variations depend on the initial eccentricity and obliquity.
The variations increase in general with the obliquity: they are nearly zero for obliquities below $40\degree$, but they can present an amplitude variation of about $1\degree$ for obliquities larger than $60^\degree$.
This can seem a modest variation, but given that the present inclination of Phobos is also $1\degree$, it corresponds to a change around $100\%$.
If the eccentricity of Mars is considered, \cite{yokoyama2005} observe that, for some initial conditions, the eccentricity and the inclination after the crossing of the resonance can even increase to about $0.08$ and $4\degree$, respectively.

\subsection{Resonance $\nu_2$}

\subsubsection{Capture probabilities}

As the obliquity of Mars is not expected to exceed $60\degree$ in the next 40~Myr, capture in the $\nu_1$ resonance is very unlikely, and so the inclination of Phobos is expected to be still close to $1\degree$ at the moment of the encounter with the $\nu_2$ resonance (Sect.~\ref{sec:incnum}).

\begin{figure}[t]
\centering
\includegraphics[scale=0.7]{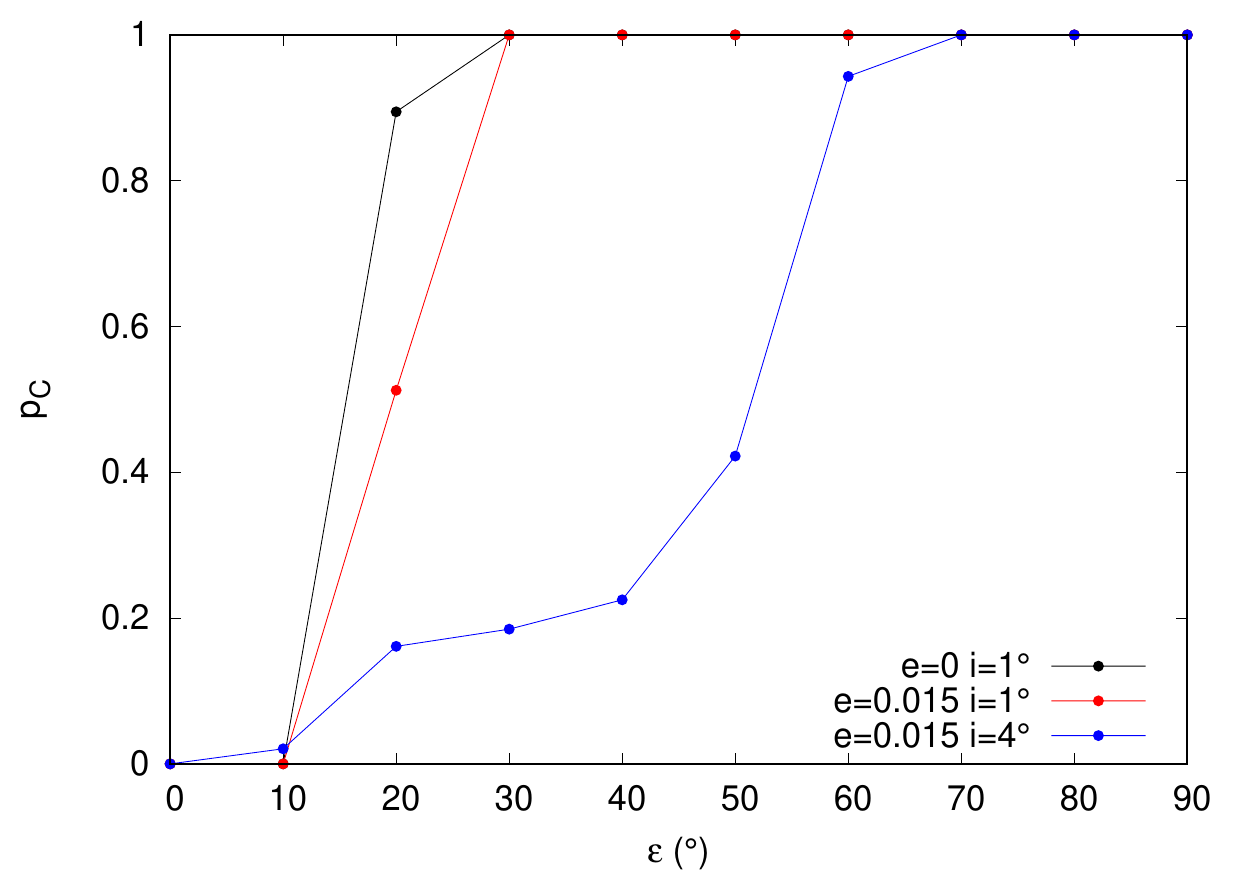}
\caption{Probabilities of capture of Phobos in the $\nu_2$ resonance with respect to the obliquity of the equator of Mars for different initial eccentricities and inclinations.\label{fig:probanum_2n1}}
\end{figure}

We compute the capture probabilities in the $\nu_2$ resonance using the present inclination of Phobos,  $i=1\degree$.
For the eccentricity, we adopt two values, $e=0$ and $0.015$, while for the obliquity we adopt values varying from $0\degree$ to $90\degree$ with a step of $10\degree$.
We start our simulations for a semi-major axis slightly larger than $a_{\nu_2}$ (Eq.\,(\ref{acprime})) and integrate the equations of motion for 250~kyr.
The capture probability statistics are shown in Fig.~\ref{fig:probanum_2n1}.

We observe that, as for the $\nu_1$ resonance, the probability of capture increases with the obliquity.
It is zero for an obliquity smaller than $10\degree$ and $100\%$ if the obliquity is higher than $30\degree$.
These results are also consistent with those observed by \cite{yokoyama2002}.
For an obliquity of $20\degree$, the probability is larger for a null initial eccentricity than for the current eccentricity, although the effect of the eccentricity is not very important.
Indeed, contrary to the $\nu_1$ resonance, which is perturbed by the nearby evection resonance, the $\nu_2$ resonance is isolated from the remaining resonances, and so the eccentricity only slightly modifies the amplitude of the libration width (Eq.\,(\ref{Ham_nu2})).

\begin{figure}[t]
\centering
\includegraphics{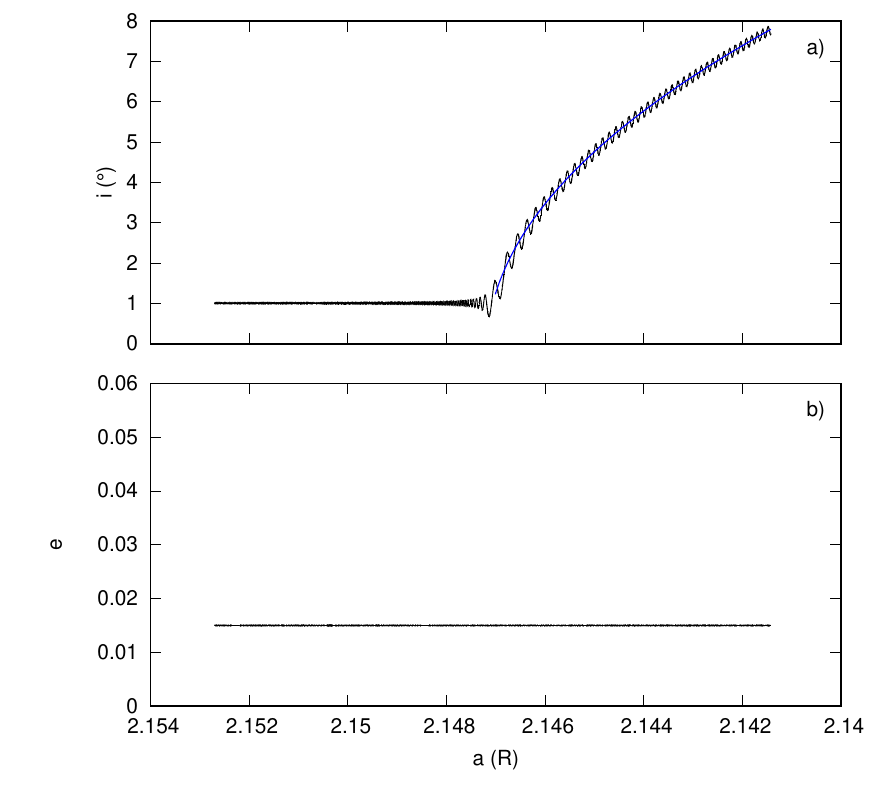}
\caption{Evolution of the inclination (a), and of the eccentricity (b) during the capture in the $\nu_2$ resonance for an obliquity of $\e=20\degree$ with the initial conditions $e=0.015$ and $\Omega=0\degree$.
The blue curve corresponds to the stable fixed point of the $\nu_2$ resonance computed with Eq. (\ref{eq:znu2}).}
\label{fig:inccap2}
\end{figure}

In Fig.~\ref{fig:inccap2}, we show an example for the evolution of the eccentricity and inclination of Phobos in the case where the capture succeeds.
We confirm that the inclination follows the equilibrium point given by expression (\ref{eq:znu2}) and that the eccentricity remains almost unchanged.

The present obliquity of Mars is about $25\degree$ and should have variations between $10\degree$ and $40\degree$ for the next 10~Myr \citep{laskar2004}.
Contrary to the $\nu_1$ resonance, we thus conclude that in the future Phobos is likely to be captured in the $\nu_2$ resonance.
As a result, we expect that the inclination of Phobos increases during the last stages of its evolution following the equilibrium point given by expression (\ref{eq_nu2}).
Therefore, although Phobos spends most of its life near the equatorial plane of Mars, we cannot rule out that the impact between Phobos and the surface of Mars will occur at high latitudes (see Fig.~\ref{fig:equilibre}).
\cite{yokoyama2005} note, however, that escape from the $\nu_2$ resonance is possible from an inclination of $33\degree$ due to interactions with several resonances.

The modification observed in the inclination during the crossing of the $\nu_1$ resonance without capture (Sect.~\ref{sec:incnum}) can impact the capture probability in the subsequent $\nu_2$ resonance.
As the capture probability decreases when the inclination increases (Sect. \ref{sec_pcap}), \cite{yokoyama2005} noted that for $i = 4\degree$ capture in the $\nu_2$ resonance is uncertain.
Therefore, in Fig. \ref{fig:probanum_2n1} we also compute the capture probability for $i=4\degree$ (blue curve) as a function of the obliquity.
Indeed, we observe that the capture probability decreases for obliquities smaller than $60\degree$, but for $20\degree < \e < 40\degree$ the capture is still possible, although with a smaller probability around $20\%$.
However, for obliquities larger than $70\degree$ the capture is certain even for $i=4\degree$.

\subsubsection{Inclination variation}

\begin{figure}[t]
\centering
\includegraphics[scale=0.7]{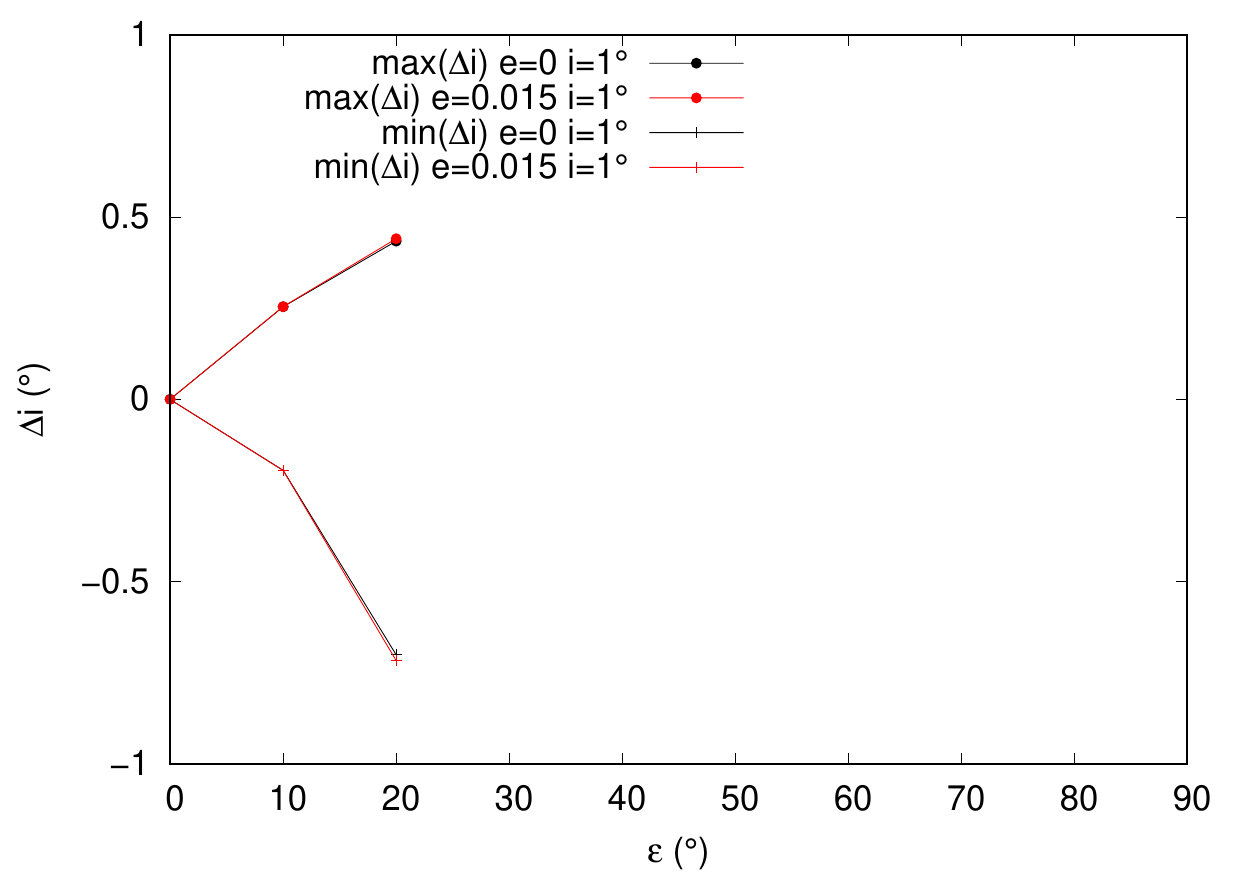}
\caption{Evolution of the maximum and minimum variations $\Delta i$ in the mean inclination owing to the encounter with the $\nu_2$ resonance in the case where the capture fails, with respect to the obliquity for different initial eccentricities.\label{fig:incnum_2n1}}
\end{figure}

In Fig.~\ref{fig:incnum_2n1}, we show the evolution of the mean inclination during the crossing of the $\nu_2$ resonance when the capture does not occur.
We only plot the results for the obliquities $0\degree$, $10\degree$, and $20\degree$, because for values of the obliquity larger than $30\degree$ the capture always occurs.
As for the $\nu_1$ resonance, the amplitude of the variation increases with the obliquity.
However, we did not observe any significant variation in the eccentricity during the crossing of the $\nu_2$ resonance.

\section{Conclusion}
\label{concdisc}

In this paper, we have revisited the secular dynamics of a satellite while perturbed by the orbital motion of its planet.
Previous studies have shown that some of these resonances can significantly modify the eccentricity (evection-like resonance) or the inclination (eviction-like resonance) of a satellite  \citep[e.g.,][]{yoder1982, touma1998, yokoyama2002, yokoyama2005}.
Eviction-like resonances are important to map, because the inclination can be used to put constrains on the formation of the satellite.
Moreover, some eviction-like resonances can be present even for the nearly zero eccentricities currently observed for the main satellites of the solar system planets.

Capture in eviction-like resonances is a possibility when the semi-major axis of the planet is decreasing.
This capture can lead to a significant increase in the inclination of the satellite.
However, for nearly zero eccentricity, capture is only possible in two of these resonances that we dubbed $\nu_1$ and $\nu_2$.
These resonances are placed close to the planets' surface, and only satellites spiraling into the planet may be affected, such as Phobos or Triton.
We analytically compute the capture probability in the $\nu_1$ resonance and compare it to numerical simulations. 
We have shown that the evolution of Phobos cannot be considered as adiabatic.

When the eccentricity of the satellite is nonzero, evection-like and additional eviction-like resonances are possible. 
Some of these resonances are close to each other, and they can interact between them. 
We have studied these interactions with the method of frequency map analysis that can be used as an alternative to surface sections, since it is easier to implement in problems with many degrees of freedom.
In the case of the $\nu_1$ resonance, we confirm that the interaction with the evection resonance $\nu_e$ introduces a chaotic region around the $\nu_1$ resonance \citep{yokoyama2002, yokoyama2005}.
In the case of Phobos, for eccentricities larger than 0.05, the evection resonance completely destroys the stability of the $\nu_1$ resonance and capture is no longer possible.
In addition, because the evection resonance occurs almost simultaneously with the $\nu_1$ resonance, a satellite that is spiraling into the planet will have its eccentricity excited by the evection resonance, which in turn decreases the chances of capture in the $\nu_1$ eviction-like resonance.

When the orbit of the satellite crosses a given secular resonance, capture is not certain.
For eviction-like resonances, the probability of capture increases with the obliquity of the planet and decreases with the inclination of the satellite.
When capture fails, the inclination of the satellite still undergoes some changes, although much smaller than in the capture scenario, and it can either increase or decrease.
In the case of the $\nu_1$ resonance, the capture probability also strongly depends on the eccentricity. However, this is a collateral effect that results from the proximity of the nearby evection resonance.

We apply our model to the study of Phobos, whose orbit will be modified in the future due to the encounters with the $\nu_1$ and $\nu_2$ eviction-like resonances.
We obtain for the capture probabilities similar results to \cite{yokoyama2002}.
The capture in the $\nu_1$ resonance is almost impossible, given the present obliquity of Mars and the present eccentricity of Phobos.
Nevertheless, the crossing of this resonance can lead to variations in the inclination between $-0.08\degree$ and $0.07\degree$.
On the contrary, capture in the following $\nu_2$ resonance is almost certain if the obliquity of Mars and the inclination of Phobos are similar to the current values, which will lead to a continuous increase in the inclination.
If the capture fails, the crossing of the $\nu_2$ resonance can lead to variations in the inclination between $-0.7\degree$ and $0.4\degree$.

We have seen that secular resonances due to stellar perturbations, such as evection-like and eviction-like resonances, can significantly influence the motion of a satellite.
The effect of planetary perturbations on satellite orbits are often considered as negligible, but they also induce secular resonances, whose crossings can lead to a one-time variation in the eccentricity or the inclination.
If the perturbations are strong enough, captures in these resonances could also be possible, and  induce more significant changes.
Future studies on the orbital evolution of satellite orbits should thus also estimate the effect from these additional secular resonances.

\begin{acknowledgements}
This work was supported by
CIDMA (UIDB/04106/2020
and UIDP/04106/2020),
CFisUC (UIDB/04564/2020 and UIDP/04564/2020),
PHOBOS (POCI-01-0145-FEDER-029932),
ENGAGE SKA (POCI-01-0145-FEDER-022217),
and GRAVITY (PTDC/FIS-AST/7002/2020),
funded by COMPETE 2020 and FCT, Portugal.
\end{acknowledgements}

\bibliographystyle{aa}
\bibliography{biblio}

\end{document}